\documentclass[twocolumn]{aastex631}

\usepackage{booktabs}
\usepackage{multirow}
\usepackage{xcolor}
\usepackage{longtable}
\newcommand{\ha}{\hbox{H$\alpha$}}
\newcommand{\hb}{\hbox{H$\beta$}}
\newcommand{\hd}{\hbox{H$\delta$}}
\newcommand{\oii}{\hbox{[O\,{\sc ii}]}}
\newcommand{\oiii}{\hbox{[O\,{\sc iii}]~$\lambda5007$}}

\newcommand{\sii}{\hbox{[S\,{\sc ii}]}}
\newcommand{\siil}{\hbox{[S\,{\sc ii}]~$\lambda\lambda6716,~6731$}}

\newcommand{\dn}{\hbox{D$_n$4000}}
\newcommand{\hda}{\hbox{$\mathrm{H\delta_A}$}}
\newcommand{\wha}{\hbox{W(H$\alpha$)}}
\newcommand{\re}{\hbox{$R_e$}}
\newcommand{\msun}{\hbox{M$_\odot$}}

\begin{document}

\title{Post-starburst Galaxies with Active Galactic Nucleus: Properties and Evolutionary Sequences}

\author[0009-0006-5571-0191]{Junjie Huang}
\affiliation{School of Astronomy and Space Science, Nanjing University, Nanjing 210023, China}
\affiliation{Key Laboratory of Modern Astronomy and Astrophysics (Nanjing University), Ministry of Education, Nanjing 210023, China}

\author[0000-0003-3226-031X]{Yanmei Chen\textsuperscript{$\star$}}
\affiliation{School of Astronomy and Space Science, Nanjing University, Nanjing 210023, China}
\affiliation{Key Laboratory of Modern Astronomy and Astrophysics (Nanjing University), Ministry of Education, Nanjing 210023, China}

\author[0000-0002-8614-6275]{Yong Shi}
\affiliation{Department of Astronomy, Westlake University, Hangzhou 310030, Zhejiang Province, China}

\author[0000-0002-8711-8970]{Cheng Li}
\affiliation{Department of Astronomy, Tsinghua University, Beijing 100084, China}

\author[0009-0001-5437-410X]{Zhuo Cheng}
\affiliation{Department of Physics, The Chinese University of Hong Kong, Sha Tin, NT, Hong Kong, China}

\author[0000-0003-0486-5178]{Ho-Hin Leung}
\affiliation{School of Physics \& Astronomy, University of St Andrews, North Haugh, St Andrews, Fife KY16 9SS, UK}
\affiliation{Institute for Astronomy, University of Edinburgh, Royal Observatory, Edinburgh EH9 3HJ, UK}

\author[0000-0002-8956-7024]{Vivienne Wild}
\affiliation{School of Physics \& Astronomy, University of St Andrews, North Haugh, St Andrews, Fife KY16 9SS, UK}

\author[0000-0002-3890-3729]{Qiusheng Gu}
\affiliation{School of Astronomy and Space Science, Nanjing University, Nanjing 210023, China}
\affiliation{Key Laboratory of Modern Astronomy and Astrophysics (Nanjing University), Ministry of Education, Nanjing 210023, China}

\author{Qihang Cheng}
\affiliation{School of Astronomy and Space Science, Nanjing University, Nanjing 210023, China}
\affiliation{Key Laboratory of Modern Astronomy and Astrophysics (Nanjing University), Ministry of Education, Nanjing 210023, China}

\author{Ying Yu\textsuperscript{$\star$}}
\affiliation{School of Astronomy and Space Science, Nanjing University, Nanjing 210023, China}
\affiliation{Key Laboratory of Modern Astronomy and Astrophysics (Nanjing University), Ministry of Education, Nanjing 210023, China}

\email{$^{\star}$E-mail: chenym@nju.edu.cn, yingyu@nju.edu.cn}

\begin{abstract}
Post-starburst (PSB) galaxies, identified by strong Balmer absorption and weak nebular emission, provide a key laboratory for studying rapid quenching. Using the final data release of the SDSS-IV MaNGA survey, we follow the traditional PSB selection criteria of \citet{Chen_2019} and develop a new method to identify regions that simultaneously exhibit PSB features and nuclear activities (AGN-PSBs). Our final sample comprises 48 AGN-PSBs, 92 central PSBs (CPSBs), 89 ring-like PSBs (RPSBs), and 828 irregular PSBs (IPSBs). We find the global and spatially resolved properties of CPSBs and RPSBs are consistent with the results of \citet{Chen_2019}. In this work, we focus on the properties of AGN-PSBs, comparing them with CPSBs, RPSBs, and control galaxies. Similar to CPSBs and RPSBs, AGN-PSBs show positive $\dn$ gradients relative to negative $\dn$ gradients of their controls, which indicates younger stellar populations in the central region than that in the outskirt. Among the three sub-types, high-mass CPSBs (H-CPSBs, with $\log(M_{*}/\msun)>9.5$) display the highest incidence of merger remnants and gas--star kinematic misalignment, consistent with a merger/interaction-dominated origin. AGN-PSBs and RPSBs, however, show lower and comparable fractions of merger remnants and gas--star kinematic misalignment, favoring less violent external mechanisms. Based on radial profiles of mass-weighted age and $V_{\rm star}/\sigma_{\rm star}$, we suggest that RPSBs can evolve into AGN-PSBs, whereas H-CPSBs likely follow a distinct evolutionary pathway. The existence of RPSBs and IPSBs also indicates that AGN feedback is not a necessary condition for the formation of PSB.
\end{abstract}


\section{Introduction} \label{sec:intro}

The galaxy population in the local universe exhibits a distinct bimodality in its color-magnitude diagram. This bimodal distribution is defined by two dominant populations: the blue cloud, comprising gas-rich galaxies undergoing active star formation, and the red sequence, consisting of gas-poor, quiescent systems, with a sparse green valley population occupying the transitional regime between them \citep{Baldry_2004,Jin_2014}. Red quiescent galaxies have undergone significant growth in both number count and total stellar mass since the epoch prior to cosmic noon \citep{Ilbert_2013,Muzzin_2013}. Such evolutionary trends imply a directional migration of galaxies from the star-forming blue cloud through the green valley to the quiescent red sequence \citep{Bell_2004,Brown_2007}. The low number density of green valley galaxies indicates that the transition process occurs on a relatively rapid timescale. Possible physical mechanisms responsible for this transition encompass both internal and external processes. Internal pathways include stellar feedback \citep{Hopkins_2014,Gatto_2017} and feedback originating from central black hole activity \citep{Goto_2006,Yesuf_2014,Baron_2017,Baron_2018}. External processes primarily involve environmental quenching \citep{Paccagnella_2018,Werle_2022,Werle_2025}, along with galaxy interactions and mergers \citep{Pawlik_2018,Wilkinson_2022}. However, the relative importance of these mechanisms remains uncertain.

To better understand how various quenching mechanisms transform star-forming galaxies into quiescent systems, it is critical to investigate galaxies undergoing rapid evolutionary transitions. Post-starburst galaxies (PSBs)—also known as E+A or K+A galaxies—represent such a population, having shifted from star-forming to quiescent states within the past $\sim$1 Gyr \citep{Dressler_1983,Poggianti_1999,Goto_2003,Goto_2005,Wild_2009,Rowlands_2018}. Marked by unusually prominent Balmer absorption lines (a signature of dominant intermediate-age A-/F-type stellar populations) and weak or absent nebular emission lines (implying a lack of hot, young O- and B-type stars), these galaxies bear witness to an abrupt cessation of star formation and thus constitute ideal astrophysical laboratories for probing the mechanisms underlying rapid galaxy quenching.

Research on post-starburst galaxies has advanced substantially alongside technological innovations in astronomical instrumentation. First identified as a distinct population in intermediate-redshift clusters via pioneering spectroscopic surveys \citep{Dressler_1983,Couch_1987}, our understanding of these systems was greatly enhanced by subsequent large-scale multi-object spectroscopic campaigns—most notably the Sloan Digital Sky Survey (SDSS; \citealt{York_2000}), which distinguishes PSBs by analyzing emission from their central 3$^{\prime\prime}$ diameter regions \citep{Goto_2003,Goto_2005,Goto_2007a,Yan_2009}. More recently, spatially resolved integral field spectroscopy (IFS) surveys—such as Mapping Nearby Galaxies at APO (MaNGA; \citealt{Bundy_2015})—have uncovered diverse types of PSB galaxies \citep{Chen_2019}, which can be classified according to the varying galactic regions where post-starburst events occur.

Based on large-scale multi-object spectroscopic surveys, extensive studies have been conducted on the physical mechanisms of the triggering and quenching of the starburst (SB) phase in PSBs. Morphological analyses reveal that PSBs are predominantly bulge-dominated systems, often with residual disk components \citep{Quintero_2004,Tran_2004,Goto_2005,Wong_2012,Maltby_2018,Pawlik_2018}, and a notable fraction exhibit disturbed morphologies or tidal features \citep{Pawlik_2016,Wilkinson_2022}. Together with constraints from stellar metallicity and abundance ratio measurements \citep{Goto_2007b,Leung_2024}, these traits strongly support short star formation timescales for PSBs. However, \cite{Sazonova_2021} demonstrated that many PSBs show only minor disturbances inconsistent with major mergers, pointing to alternative triggers such as internal instabilities or obscured interactions. Observational studies have also identified environmental effects on PSB formation. Surveys of massive galaxy clusters show a higher PSB fraction than in the field \citep{Dressler_1999,Poggianti_1999,Tran_2003,Tran_2004,Linden_2010,Socolovsky_2018,Paccagnella_2018}, implying that environmental processes (e.g., ram-pressure stripping, galaxy harassment) may trigger the abrupt truncation of star formation activity \citep{Werle_2022,Werle_2025}. Notably, most local PSBs are field systems with no obvious correlation with local density \citep{Quintero_2004,Balogh_2005,Goto_2005,Hogg_2006,Yan_2009,Pawlik_2018}. Furthermore, the role of active galactic nucleus (AGN) feedback in PSB quenching remains contested: while some studies report clear evidence of AGN-driven quenching via gas expulsion \citep{Goto_2006,Yesuf_2014,D'Eugenio_2024}, others argue that low-$z$ AGNs lack sufficient energy to radiatively eject cold gas \citep{Lanz_2022}. Complementing observational results, EAGLE simulations further reveal diverse evolutionary pathways for local PSBs \citep{Pawlik_2019}. These conflicting lines of evidence highlight that further studies are essential to unravel the origin and evolution of these intriguing galaxies.

The MaNGA IFS survey has yielded spatially resolved measurements of the kinematics, stellar populations as well as metallicities of PSB galaxies. On the one hand, these observational results serve as critical empirical constraints for theoretical models, thereby advancing our understanding of the formation mechanisms of PSBs. On the other hand, they have also identified a diversity of PSB sub-types, which not only adds complexity to related research but also highlights the necessity for more in-depth investigations into their evolutionary pathways. \cite{Chen_2019} analysed galaxy spectra of individual spatial pixels (spaxels) from the MaNGA survey and found 360 galaxies with PSB regions. These 360 galaxies are classified into three types: 31 galaxies with central contiguous post-starburst regions (CPSB), 37 galaxies with off-centre ring-like post-starburst regions (RPSB) and the remaining 292 galaxies with irregular post-starburst regions (IPSB). The discovery of IPSB galaxies indicates that the rapid quenching for PSB signatures need not be galaxy-wide but can occur locally. While \cite{Chen_2019} found that both CPSBs and RPSBs host young stellar populations in their central regions, the significant discrepancies in the average radial profiles of stellar velocity-dispersion ratio ($V_{\rm star}/\sigma_{\rm star}$) and mass-weighted age led them to propose that these two sub-types primarily originate from distinct formation channels. Later studies by \cite{Cheng_2024} observed similar outside-in quenching patterns in CPSBs and RPSBs, suggesting these two sub-types may be different stages of the same event. While \cite{Chen_2019} and \cite{Cheng_2024} relied on $\dn$ as a stellar population indicator, accurate star formation histories (SFH) of both rapidly quenched PSB and non-PSB regions in both CPSBs \citep{Leung_2024} and RPSBs \citep{Leung_2025} have been obtained through Bayesian full spectral fitting. Combined with chemical evolution measurements of both PSB types, \cite{Leung_2025} discussed in detail the possible evolutionary connection and likely origins of CPSBs and RPSBs.

Almost all studies reviewed herein—particularly those based on IFS datasets—have employed traditional selection methods for identifying PSBs. The classical approach selects systems with strong Balmer absorption lines and weak or absent nebular emission lines \citep{Quintero_2004,Tran_2004,Blake_2004,Yang_2008,Pawlik_2018,Chen_2019}. However, this criterion can miss galaxies in which the quenching process produces substantial emission, for example from shocks driven by stellar winds, AGNs, or low-ionization nuclear emission-line regions (LINERs). PSBs hosting concurrent AGN activity therefore provide key laboratories for testing the physical connection between AGN activity and the PSB phenomenon.

To mitigate this limitation, several alternative selection techniques have been developed. For example, \citet{Goto_2006} selected galaxies exhibiting strong $\hd$ absorption while simultaneously applying Baldwin–Phillips–Terlevich (BPT) classifications \citep{Baldwin_1981,Veilleux_1987,Kauffmann_2003,Kewley_2001,Kewley_2006}. \citet{Alatalo_2016} introduced the class of shocked post-starburst galaxies (SPOGs), which permits ionized emission consistent with shocks, AGNs, and LINERs. These approaches generally require strong $\hd$ absorption while allowing emission lines powered by shocks, AGNs, or LINER-like mechanisms, rather than restricting the emission solely to star formation. A comprehensive review of PSB selection criteria is provided by \citet{French_2021}.

In this work, we adopt a composite selection that combines prominent Balmer absorption features with BPT-identified AGN activity to select PSB galaxies/regions with AGN-powered nebular emission lines (AGN-PSBs for short). Applying this method to the 10,010 unique galaxies in the final data release of the MaNGA survey, together with the traditional selection approach of \citet{Chen_2019}, we construct the largest spatially resolved PSB sample to date. We then perform a systematic analysis of these samples to investigate the properties of AGN-PSBs, comparing them with CPSBs, RPSBs, as well as control galaxies.

The paper is organized as follows: Section 2 describes the MaNGA data and sample selection criteria. Section 3 presents global and spatially resolved properties of PSB samples and their relevant controls, as well as their comparison. Section 4 discusses the observational results, followed by a summary in Section 5.

\section{Data} \label{sec:data}
\subsection{The MaNGA Survey} \label{sec:manga}

MaNGA is one of three core programs of the fourth-generation Sloan Digital Sky 
Survey (SDSS-IV; \citealt{Bundy_2015,Drory_2015,Law_2015,Law_2016,Yan_2016a,Yan_2016b,Blanton_2017}), implemented using the 2.5m Sloan Foundation Telescope \citep{Gunn_2006} at Apache Point Observatory (APO). It employs 
dithered observations with 17 fiber-bundle integral field units (IFUs), which cover five distinct sizes ranging from 19 to 127 fibers, corresponding to a sky-projected diameter of 12.5$-$32.5$^{\prime\prime}$. Two dual-channel BOSS spectrographs \citep{Smee_2013} provide simultaneous spectral coverage over 3600$-$10300\AA\ at a median spectral resolution of $R\sim 2000$. The final data release of MaNGA survey includes 10,010 unique galaxies with a redshift range of $0.01 < z < 0.15$ and a roughly flat stellar mass distribution between $10^9 \msun$ -- $10^{11} \msun$ \citep{Wake_2017}. 2/3 of the galaxy sample are observed out to $\sim$1.5 effective radii ($\re$) and the other 1/3 to $\sim$2.5$\re$. With a typical integration time of 3 hours, MaNGA reaches a $r-$band signal-to-noise ratio (S/N) of $4\sim 8$ per fiber at a surface brightness of 23 ABmag arcsec$^{-2}$, which is typical for the outskirts of MaNGA galaxies \citep{Yan_2016a}. The $2^{\prime\prime}$ fiber diameter corresponds to a physical scale of $\sim$ 1~kpc at the peak redshift ($z \sim$ 0.03) of the MaNGA sample.

\subsection{Data Analysis}\label{sec:dap}
The MaNGA Data Reduction Pipeline (DRP; \citealt{Law_2016}) processes raw observational data to generate flux-calibrated and sky-subtracted 3D spectra for each target galaxy. Complementary to the DRP, the MaNGA Data Analysis Pipeline (DAP; \citealt{Westfall_2019}) is the survey-led software package designed to analyze the DRP-produced spectra, yielding high-level, science-ready data products. The DAP executes two distinct full-spectrum fitting procedures, both using the penalized pixel-fitting (pPXF) algorithm \citep{Cappellari_2004} but employing different stellar templates. For the determination of stellar kinematics, the pipeline utilizes templates from the MILESHC library \citep{Sanchez_2006}; in contrast, the stellar continuum fitting within the emission-line analysis module relies on templates from MASTARSSP. Notably, MILESHC and MASTARSSP are both stellar-template libraries constructed via hierarchical clustering—MILESHC from the MILES library, and MASTARSSP from the MaStar library \citep{Yan_2019}. Key outputs of the DAP include quantitative estimates of stellar absorption lines, as well as measurements of 21 major nebular emission lines spanning the MaNGA wavelength coverage.

For this work, the projected stellar rotation velocity ($v_{\rm star}$), stellar velocity dispersion ($\sigma_{\rm star}$), rotation velocity ($v_{\rm gas}$) and velocity dispersion ($\sigma_{\rm gas}$) of ionized gas, spectral indices ($\dn$ \& $\hda$) as well as fluxes and equivalent widths of nebular emission lines are extracted from the DAP products named ``MAPS-SPX-MILESHC-MASTARSSP'', which provides pixel-wise analysis results for individual spaxels. It is noted that the emission line fluxes and equivalent widths are presented after subtraction of the underlying stellar continuum absorption. The index $\dn$ represents the strength of the 4000\AA\ break defined as the flux density ratio between two narrow continuum bands 3850$\sim$3950 and 4000$\sim$4100\AA\ \citep{Bruzual_1983}. The Lick Index $\hda$ quantifies the equivalent width of $\hd$ absorption feature in the bandpass 4083$-$4122\AA\ with continuum bandpasses of 4041.6$-$4079.75\AA\ and 4128.5$-$4161.0\AA\ \citep{Worthey_1994,Worthey_1997}.

The derived global galaxy parameters utilized in this work include the global stellar mass ($M_*$), $\re$, $g-r$ color, S$\acute{\rm e}$rsic index, as well as luminosity-weighted and mass-weighted stellar population ages. The stellar masses we use are derived from Spectral Energy Distribution (SED) fitting, based on the K-corrected S$\acute{\rm e}$rsic fluxes from the MaNGA DRP catalog\footnote{\url{https://data.sdss.org/datamodel/files/MANGA_SPECTRO_REDUX/DRPVER/drpall.html}} and adjusted to a Hubble constant of $h = 0.7$. We also use the Elliptical Petrosian 50\% light radius in SDSS r-band as $\re$. The S$\acute{\rm e}$rsic index is taken from the MaNGA PyMorph photometric catalog \citep{Fischer_2019}, adopting the value from the single-component S$\acute{\rm e}$rsic fit. The color index, luminosity-weighted and mass-weighted stellar ages, are obtained from the MaNGA-Pipe3D value-added catalog \citep{Sanchez_2016a,Sanchez_2016b}. 

\subsection{Sample Selection} \label{sec:psb sample}
Multiple approaches have been proposed for the selection of PSBs. Traditionally, PSBs are identified by their distinctive spectral signatures: strong Balmer absorption lines (a tracer of recent starburst activity) and weak or absent nebular emission lines (e.g., $\ha$ and/or $\oii$), which indicate the lack of ongoing star formation. However, the strict constraints imposed on nebular emission lines introduce two key biases: on the one hand, it biases the selection against galaxies hosting shocks and type 2 AGNs; on the other hand, it excludes PSBs that are not (yet) fully quenched \citep{Yan_2006, Wild_2007, Wild_2009, Kocevski_2011, Yesuf_2014, Alatalo_2014}. By revising the emission-absorption line diagnostic threshold, \cite{Chen_2019} proposed a selection criterion to identify galaxies with $\ha$ nebular emission strength that are lower than expected from the strength of their Balmer absorption lines. In the present work, we closely adopt the methodology described in \cite{Chen_2019} to select both traditional PSBs and those undergoing quenching. Furthermore, we develop a supplementary selection criterion to include PSBs hosting central AGNs or shocks, thereby mitigating the remaining limitation of the original method.

\subsubsection{Selection of Traditional PSBs} \label{sec:typical PSB}

\begin{figure*}[htbp]
    \centering
    \includegraphics[width=0.6\textwidth]{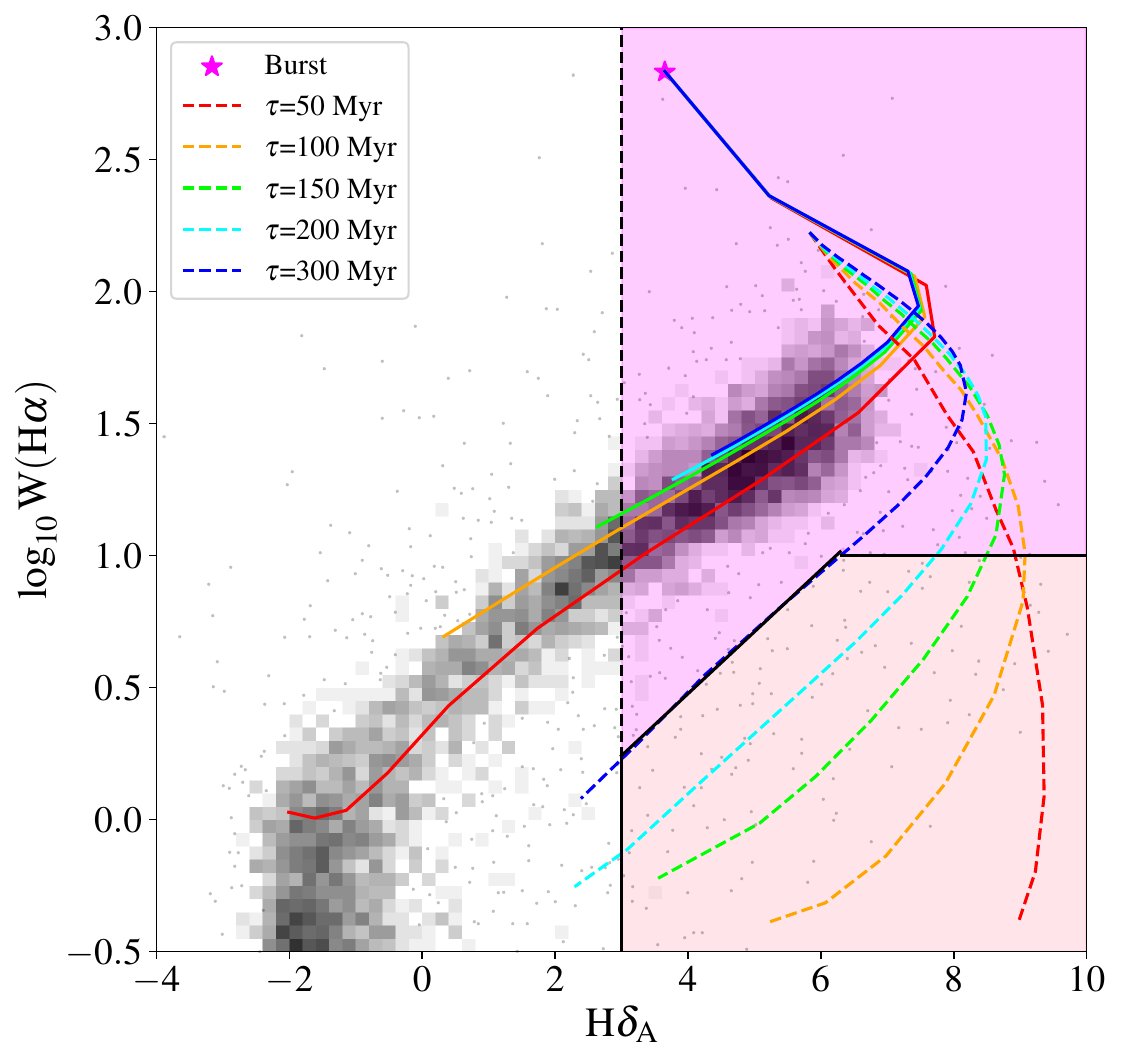}
    \caption{The relationship between the $\hd$ absorption line and the $\ha$ emission line equivalent width for a sample of galaxies from the MaNGA DAP catalog, overlaid with the toy model evolutionary tracks. The solid tracks correspond exponentially declining SFHs with e-folding timescales ranging from $\tau$ = 0.5 Gyr (red) to $\tau$ = 5 Gyr (blue). The dashed tracks correspond to models featuring an extra starburst at 6.5 Gyr, followed by a star formation truncation with the e-folding timescales specified in the legend. All the models have a common youngest point marked by the magenta star, and they evolve towards the bottom-left corner over time. The pink-shaded area denotes the traditional PSB region \citep{Chen_2019}, while the magenta-shaded region corresponds to the AGN-PSB region.}
    \label{fig:1}
\end{figure*}

Fig.~\ref{fig:1} shows the correlation between $\hda$ and the equivalent width of $\ha$ ($\wha$) from the MaNGA DAP catalog\footnote{\url{https://data.sdss.org/datamodel/files/MANGA_SPECTRO_ANALYSIS/DRPVER/DAPVER/dapall.html}}, overlaid with evolutionary tracks from toy models. For reliable $\hda$ measurements, we require the spectra to have median S/N $>$ 10 per pixel for the following analysis. A spaxel is classified as a traditional PSB if it satisfies three criteria: $\wha<10$\AA, $\hda>3$\AA\ and $\log\wha<0.23\times\hda- 0.46$ (see the pink-shaded area in Fig.~\ref{fig:1}). These selection criteria are identical to those defined in \cite{Chen_2019}.


\begin{figure*}[htbp]
    \centering
    \includegraphics[width=1\textwidth]{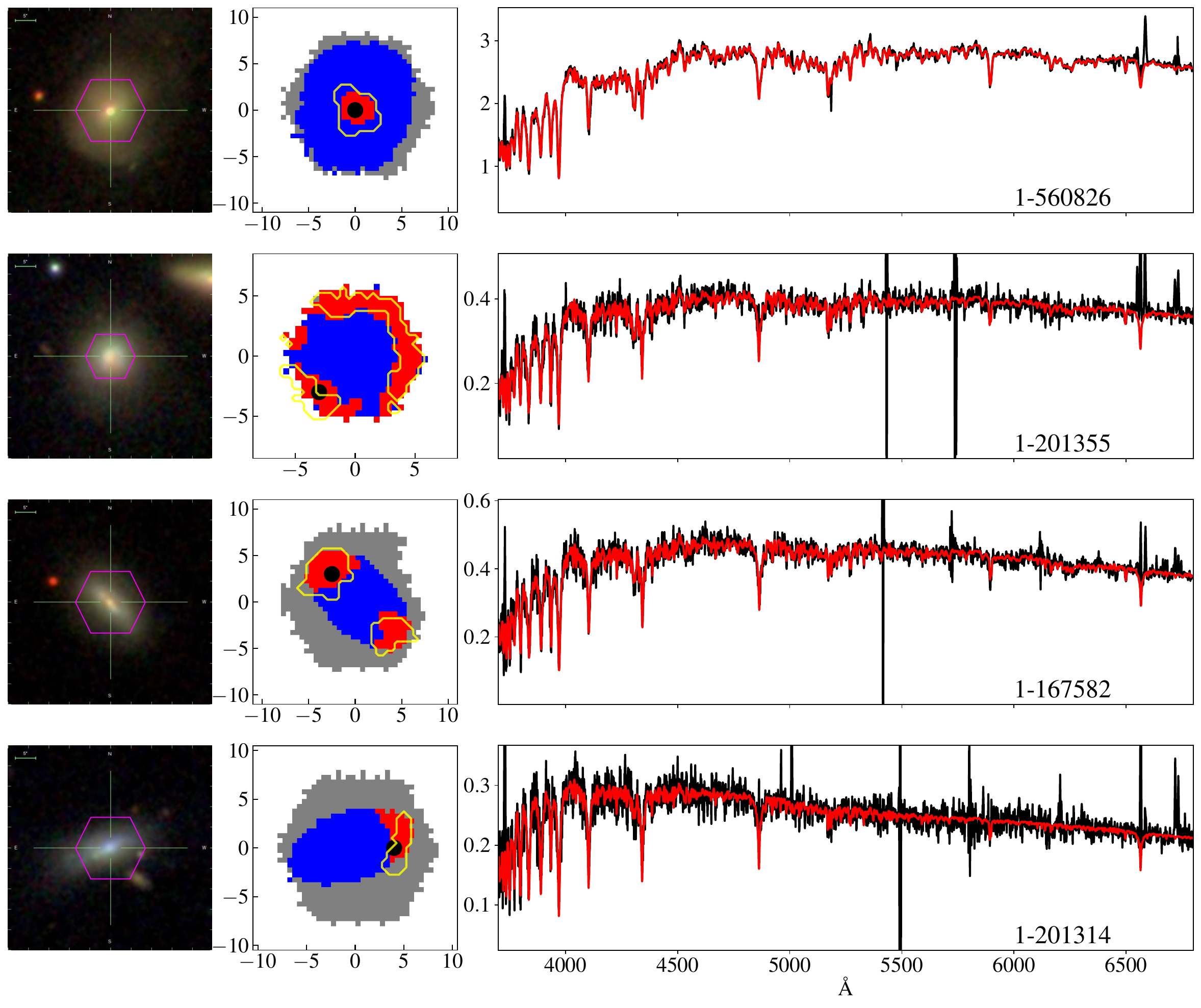}
    \caption{Examples of four PSBs in the MaNGA survey. The left panel shows the SDSS $g,r,i$-image. The middle panel colors PSB spaxels in red, non-PSB spaxels with median spectral S/N per pixel $>$ 10 in blue, and spaxels with lower-S/N in grey; overlaid yellow lines denote the PSB regions identified by \cite{Cheng_2024}. The right-hand panel shows the spectrum at the location indicated by the black dot in the middle panel, with the observed spectrum shown in black and the best-fit stellar continuum model derived from the MaNGA DAP in red. The MaNGA ID of the example galaxy is labeled at the bottom right of each row. The figure is organized from top to bottom as follows: a CPSB, a face-on RPSB, an edge-on RPSB, and an IPSB system. As expected, the PSB regions selected by us basically overlap with \citet{Cheng_2024}.}
    \label{fig:2}
\end{figure*}

After excluding contaminants (foreground stars and background galaxies) which affect the continuum fitting, we identify 1061 galaxies from an initial sample of 10,010, each containing more than 6 contiguous spaxels that meet the selection criteria outlined above. As established in \cite{Chen_2019}, PSB galaxies can be categorized into three distinct groups according to the spatial location and morphology of their PSB regions: (1) galaxies with central PSB regions (CPSBs), (2) galaxies with ring-like PSB regions (RPSBs), and (3) galaxies with irregular PSB regions (IPSBs). To ensure the reliability of PSB classification, we remove ongoing merging systems which introduce ambiguities in the identification of these diagnostic structural features; this step reduces the sample size to 1014 galaxies. Following the classification scheme of \cite{Chen_2019}, our final sample comprises 92 CPSBs, 94 RPSBs, and 828 IPSBs. Fig.~\ref{fig:2} shows four example galaxies of these three PSB sub-types. 


Given the irregular spatial distribution of PSB regions in IPSBs---likely driven by bright star clusters, intermediate-age cluster complexes, or other localized physical processes \citep{Chen_2019}---we exclude IPSBs from subsequent analyses. We cross-match our CPSB/RPSB samples against the PSB catalog compiled by \cite{Cheng_2024}, which draws from the same parent sample. The catalog construction workflow of \cite{Cheng_2024} is as follows: spaxel exhibiting the highest $\hda$ ($>$3\AA) is selected as the center of a potential ``$\hd$-strong region''. Adjacent spaxels with $\hda$ ($>$3\AA) within a distance of $r_{p}^{\max}$ (= max\{1.5$^{\prime\prime}$, 500 pc\}) from this center are then incorporated into this region. Each identified ``$\hd$-strong region'' is removed from the map, and the process is repeated on the remaining spaxels until all spaxels are either assigned to a region or discarded. For each region, spectra were then co-added to increase the S/N, with subsequent measurements of the $\hda$ and $\wha$ performed on the combined spectra. Based on the $\hda$ and $\wha$ values obtained from the stacked spectra, they further judged whether the region was a PSB region. Of the 92 CPSBs in our sample, 79 overlap as CPSBs in \citet{Cheng_2024}; of the 13 non-overlapping sources, 4 and 2 were classified as RPSBs and IPSBs respectively in their work, with the remaining 7 excluded from their selection. Among the 94 RPSBs, 70 overlap as RPSBs in \citet{Cheng_2024}; for the 24 non-overlapping RPSBs, 20 were categorized as IPSBs in their dataset, with the remaining 4 omitted from their sample. CPSBs and RPSBs identified exclusively in the catalog of \citet{Cheng_2024} are predominantly merging systems or targets with unreliable DAP fits.

\subsubsection{Selection of PSBs hosting shocks or central AGN} \label{sec:PSB with AGNs}

While the methodology outlined in the preceding section efficiently identifies PSBs, the cut applied to $\ha$ emission line strength overlooks PSB systems where AGNs and/or shocks produce prominent emission lines \citep{Quintero_2004,Tran_2004,Blake_2004,Yang_2008,Pawlik_2018,Chen_2019}. However, galaxies that simultaneously exhibit the characteristics of both PSB (strong Balmer absorption lines) and AGNs are ideal laboratories for investigating the connection between black hole activity and quenching of star formation. We therefore propose the following criteria to identify AGN-PSB galaxies/regions. Again, we only include spaxels with median spectral S/N per pixel $>$ 10 for the analysis. A spaxel located in the magenta-shaded area in Fig.~\ref{fig:1} is selected as AGN-PSB if it satisfies two conditions: \hda$>$3\AA\ and is classified as AGN in the $\sii$-BPT diagram. To ensure robust diagnostic results, all four emission lines ($\ha$, $\hb$, $\oiii$ and $\siil$) applied in the BPT diagram must have S/N $>$ 3.

\begin{figure*}[htbp]
    \centering
    \includegraphics[width=1\textwidth]{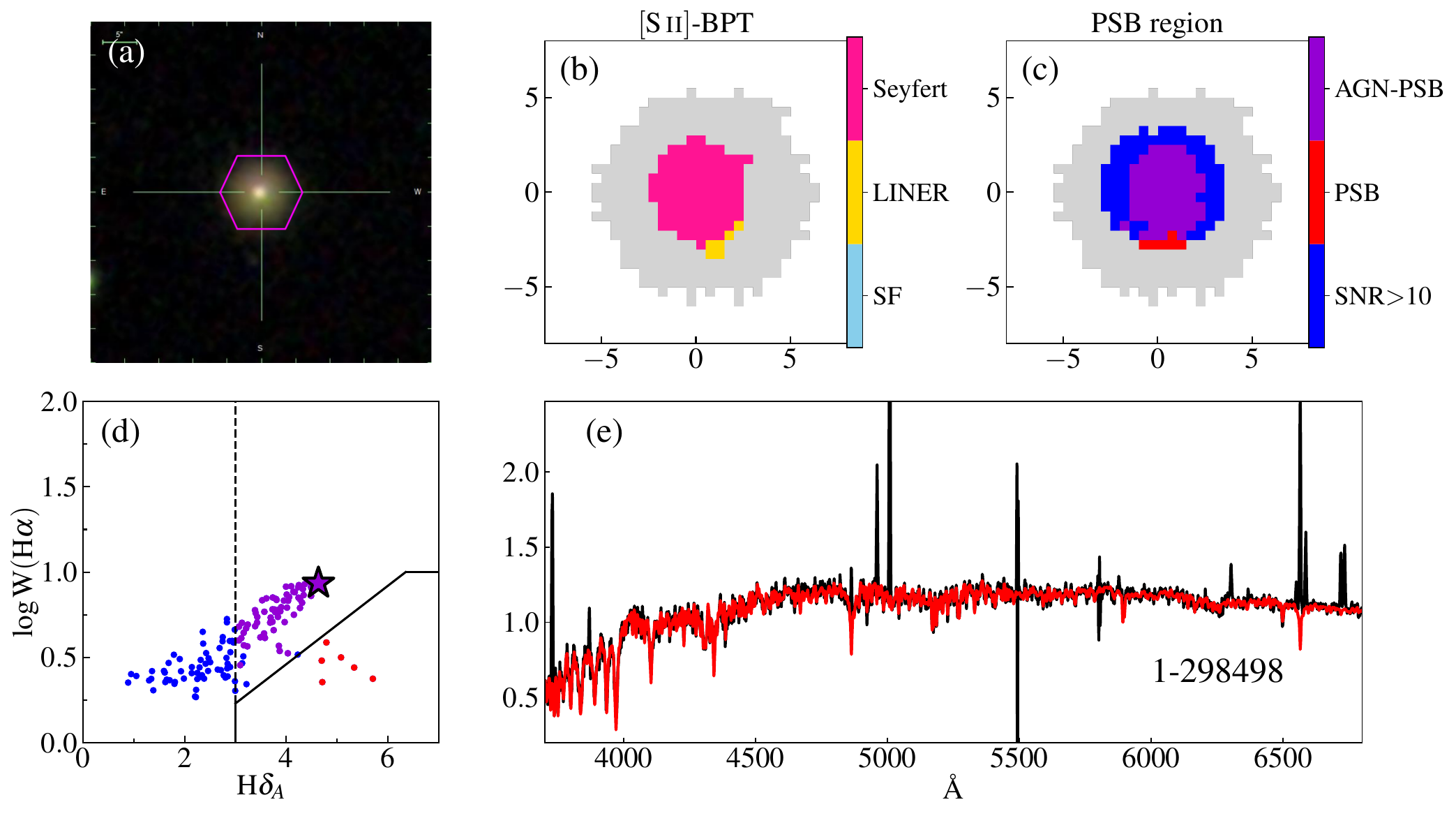}
    \caption{An example of AGN-PSB galaxy. Panel (a) displays the SDSS $g,r,i$-image. Panel (b) displays the spatially resolved $\sii$-BPT diagram, with Seyfert spaxels in pink, LINER spaxels in yellow, SF spaxels in blue. Panel (c) highlights the spatial distribution of spaxels: AGN-PSB spaxels in purple, traditional PSB spaxels in red, spaxels of median spectral S/N $>$ 10 per pixel in blue, and lower S/N spaxels in gray. Panel (d) presents the $\wha$-$\hda$ correlation, where the central spaxel is marked by a star, AGN-PSB spaxels in purple, traditional PSB spaxels in red, and other spaxels with median spectral S/N per pixel $>$ 10 in blue. Panel (e) shows the central spectrum in black, with the best-fit stellar continuum model from the MaNGA DAP overplotted in red. MaNGA ID of the galaxy is labeled at the bottom right.}
    \label{fig:3}
\end{figure*}

From 10,010 galaxies we select 309 AGN-PSB candidates with more than 6 contiguous spaxels that meet our selection criteria. We then visually exclude ongoing mergers (which preclude reliable radial gradient analyses), broad-line AGN hosts (for which emission/absorption-line measurements provided by DAP are unreliable), and systems exhibiting strong $\hda$ absorption with AGN-like line ratios at galactic outskirts rather than in the central region (in these galaxies, the line ratios are tend to be contaminated by diffuse ionized gas), leaving a final sample of 48 AGN-PSB galaxies, with PSB regions confined to galaxy centers. Fig.~\ref{fig:3} presents an AGN-PSB example. Five of the 48 AGN-PSB galaxies were previously classified as RPSBs in the last section. We reclassify these objects as AGN-PSBs herein, yielding a final sample of 89 pure RPSBs for subsequent analyses. Tab.~\ref{tab:1} summarizes the number of PSBs for each type.

\begin{longtable}{lcccc}
\caption{The Number of PSBs for Each Type}
\label{tab:1}\\
\toprule
Type     & CPSB & RPSB & AGN-PSB & IPSB \\
\midrule
\endfirsthead

\toprule
Type     & CPSB & RPSB & AGN-PSB & IPSB \\
\midrule
\endhead

Number   & 92   & 89   & 48      & 828  \\
\bottomrule
\end{longtable}

\subsection{Selection of Control Samples} 

To quantify the differences between PSBs and the broader galaxy population, we construct control samples by matching each PSB galaxy to its non-PSB counterpart based on two key parameters: global stellar mass and $\dn$. The global $\dn$ is measured from the stacked spectra across the full spatial extent of the MaNGA bundle. The motivation for selecting these parameters is as follows: (i) constraining the control galaxies to have similar stellar masses is essential, given the strong dependence of stellar population properties on stellar mass, (ii) matching the global $\dn$ ensures that the control samples have comparable light-weighted stellar ages averaged over the past few gigayears, corresponding to epochs prior to the onset of PSB-related events. Although this matching is not perfect---since $\dn$ can increase following the shutdown of star formation---it provides a practical means of controlling for stellar population age without invoking extensive, model-dependent spectral fitting, which is beyond the scope of this work.

For AGN-PSBs, we further require the line ratios of the central 1~kpc of the control galaxies to fall within the same ionization regime in the $\sii$-BPT diagram \citep{Baldwin_1981, Veilleux_1987} as those of AGN-PSBs. This criterion ensures that the central regions of both AGN-PSBs and their control sample are ionized by central black hole activities, thereby minimizing the influence of black hole accretion processes on the comparison results. Following the methodology of \cite{Alban_2023}, we stack the spectra of spaxels within a central circular aperture of 1~kpc radius for all 7,503 emission-line galaxies in the final MaNGA data release, which have $\ha$ emission-line S/N greater than 3 for at least 10\% of spaxels within 1.5$\re$. We employ weighted averaging: if a spaxel partially contributes to the circular aperture, its weight is determined by the fraction of its area enclosed within the aperture. \citet{Alban_2023} investigated the impact of aperture sizes on AGN selection, demonstrating that the central 1 kpc circular region is a suitable choice for classifying the central ionization state.

\begin{figure}[htbp]
    \centering
    \includegraphics[width=0.4\textwidth]{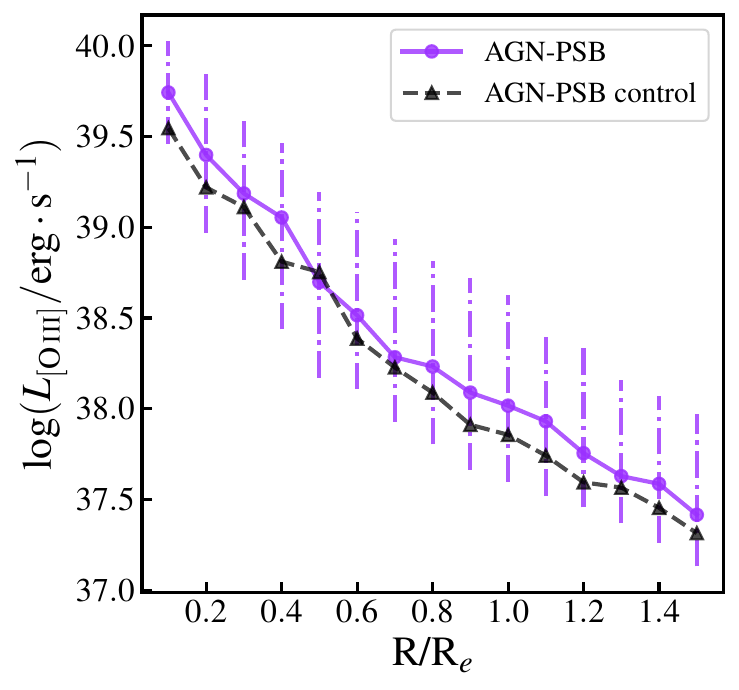}
    \caption{The median radial profiles of log($L_{\mathrm{[O\, III],\,1kpc}}$) for AGN-PSBs (purple solid line) and their controls (black dashed line). The error bars show the 30th to 70th percentile of the distribution for the AGN-PSBs.}
    \label{fig:4}
\end{figure}

We correct the $\oiii$ luminosity for dust attenuation using the Balmer decrement $\ha/\hb$, assuming Case~B recombination, and adopting the attenuation curve from \citet{Calzetti_2001}. The dust-corrected $\oiii$ luminosity is used in our subsequent analysis. We adopt the $\oiii$ luminosity within the central 1~kpc ($L_{\mathrm{[O\,III],\,1kpc}}$) as a tracer of AGN activity strength, which is measured from the stacked central-aperture spectrum. The control sample for AGN-PSBs is then constructed by matching each AGN-PSB to its nearest non-PSB AGN host based on three parameters: global stellar mass, $\dn$, and $L_{\mathrm{[O\,III],\,1kpc}}$. Fig.~\ref{fig:4} compares the median radial profiles of $\log(L_{\mathrm{[O\,III]}})$ for AGN-PSBs (purple solid line) and their controls (black dashed line). These profiles exhibit nearly identical radial gradients, confirming that the control galaxies possess AGN activity levels comparable to those of the AGN-PSBs. This matching approach ensures that we can isolate differences between AGN-PSBs and their control sample that are not driven by nuclear activity in subsequent analyses.

\section{Results} \label{sec:results}

Tables~\ref{tab:agnpsb}, \ref{tab:cpsb}, and \ref{tab:rpsb} list the samples of AGN-PSB, CPSB and RPSB galaxies, respectively, along with the relevant parameters employed in this work.

\begin{figure*}[htbp]
    \centering
    \includegraphics[width=0.8\textwidth]{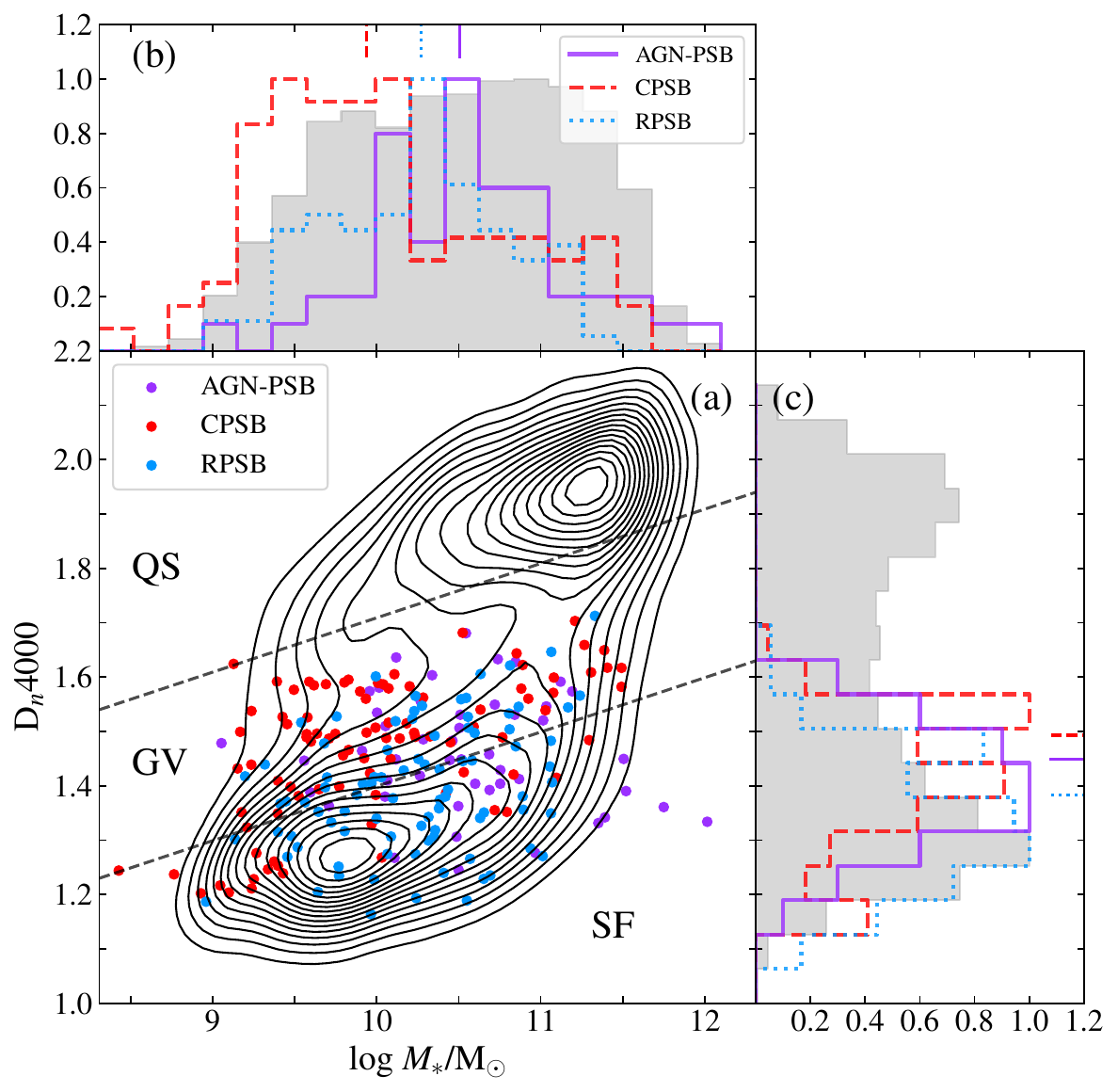}
    \caption{The distribution of three PSB sub-types on the global $\dn$–stellar mass relation. The background contours represent the full MaNGA sample, and the two dashed lines divide galaxies into star-forming (SF), green-valley (GV), and quiescent-sequence (QS) galaxies. The lower dashed line approximates the upper boundary of the SF main sequence (at the $\sim$1$\sigma$ scatter level), while the upper dashed line marks the lower boundary of the QS. AGN-PSBs, CPSBs, and RPSBs are overlaid in purple, red, and blue, respectively. The top and right histograms show distributions of stellar mass and global $\dn$, respectively. The gray shaded histograms represent the full MaNGA sample. Each distribution is normalized to its peak value for clear comparison, and the vertical lines at the top of these panels mark the median of these distributions.}
    \label{fig:5}
\end{figure*}

Fig.~\ref{fig:5} shows the distribution of three PSB sub-types on the global $\dn$–stellar mass relation. We find that, 56\% of RPSBs reside in the blue cloud, compared to 23\% for CPSBs and 37\% for AGN-PSBs. All other PSBs are located in the green valley region, between the red sequence and blue cloud.


Panel (b) shows the distribution of stellar mass for AGN-PSBs, CPSBs, and RPSBs. Each distribution is normalized to its peak value for clear comparison. AGN-PSB galaxies tend to have the highest median stellar mass ($\log(M_{*}/\msun)=10.2$), while CPSBs exhibit the lowest median stellar mass ($\log(M_{*}/\msun)=9.6$) but the broadest mass coverage. The median $\log(M_{*}/\msun)$ of RPSB galaxies is 10.0, falling between those of CPSBs and AGN-PSBs. Notably, a large fraction of CPSBs have low stellar masses ($\log(M_{*}/\msun)<9.5$); this population will be discussed in detail in section~\ref{sec:psb fraction}. Panel (c) displays the distributions of global $\dn$. RPSBs have the youngest stellar populations with a median $\dn$ of 1.38, while CPSBs and AGN-PSBs have median global $\dn$ of 1.49 and 1.45, respectively.

These distinct distributions of stellar mass and stellar population age may suggest potential differences in evolutionary scenarios and quenching mechanisms among the three PSB sub-types. In the following sections, we investigate the properties and evolutionary connections among AGN-PSBs, CPSBs and RPSBs. We first examine the PSB fraction as a function of galaxy properties, followed by a statistical analysis of their kinematics and morphology. We ultimately compare the global and spatially resolved properties across the three PSB sub-types.

\subsection{Fraction of PSBs} \label{sec:psb fraction}

\begin{figure*}[htbp]
    \centering
    \includegraphics[width=0.9\textwidth]{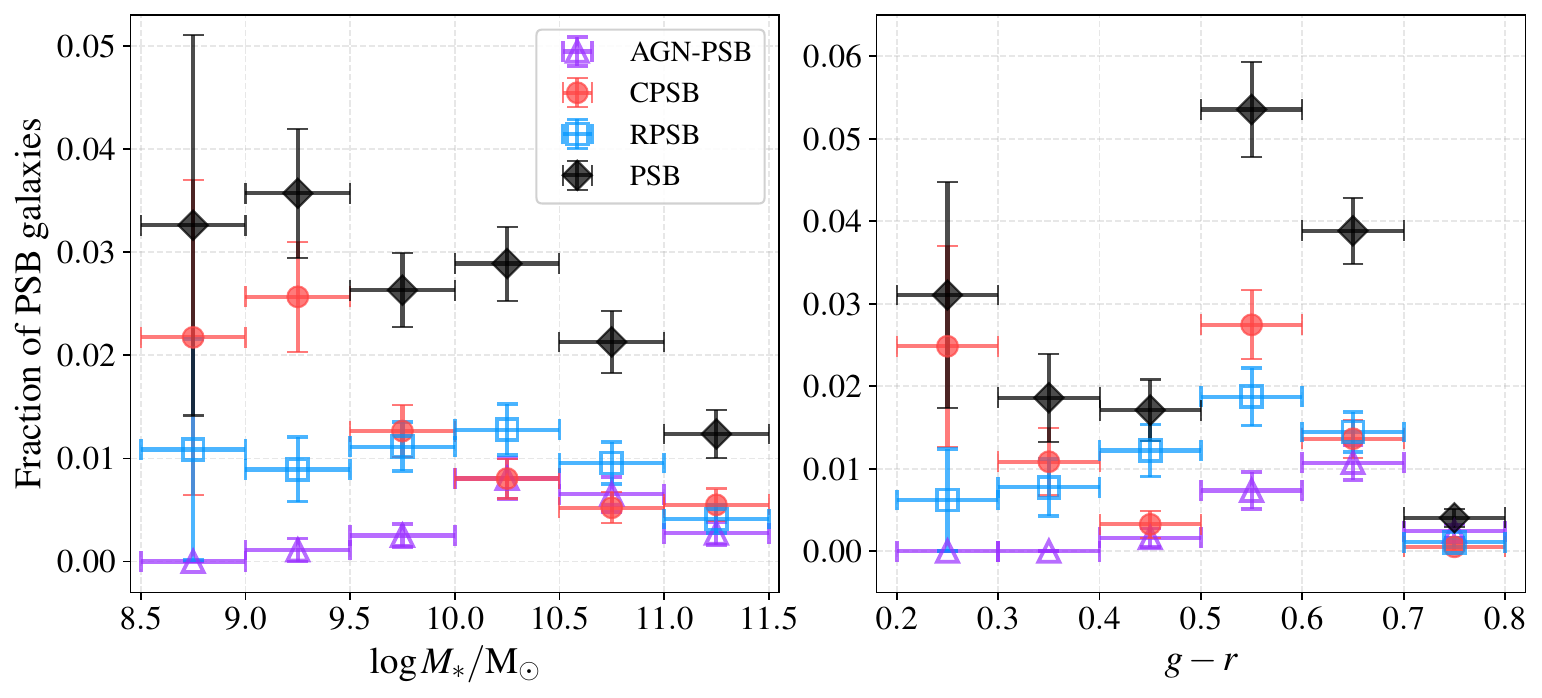}
    \caption{Fraction of CPSBs (red circles), RPSBs (blue squares), AGN-PSBs (purple triangles) and the total PSB population (AGN-PSBs$+$CPSBs$+$RPSBs) (black diamonds) as a function of stellar mass and $g-r$. The error on the x-axis is defined by the parameter binsize, while the error on the y-axis is estimated through bootstrap resampling.}
    \label{fig:6}
\end{figure*}

In this section, we study the fraction of PSBs as a function of two key parameters: stellar mass and $g-r$ color. The results are shown in Fig.~\ref{fig:6}. In the left panel, the fraction of the total PSB population decreases with increasing stellar masses. Turning to specific classes of PSBs, we find that AGN-PSBs reach its maximum around $9.5 < \log(M_{*}/\msun) < 10.5$, and declines toward both ends of the mass distribution. RPSBs remain approximately constant at $\sim$1\% within the range of $8.5 < \log(M_{*}/\msun) < 10.5$ and decrease toward higher stellar masses. In contrast, CPSBs decrease monotonically with stellar mass, falling from $\sim$2\% at $\log(M_{*}/\msun) \sim 9$ to $\sim$1\% at $\log(M_{*}/\msun) \sim 10$, which dominate the low-mass end of the total PSB population. In the right panel, the fraction of total PSB population shows bimodal distribution, with peaks at $g-r\sim$0.2--0.3 and $\sim$0.5--0.6, respectively. Among these, AGN-PSBs and RPSBs peak at $g-r\sim$0.6--0.7 and $\sim$0.5--0.6, respectively, and they decrease toward both the red and the blue ends. However, CPSBs display a bimodal distribution in $g-r$, one peak located at $g-r \sim$0.2--0.3 while the other located at $g-r \sim$0.5--0.6, making them the dominant contributors to the total PSB population at the blue-color end.


\begin{figure*}[htbp]
    \centering
    \includegraphics[width=1\textwidth]{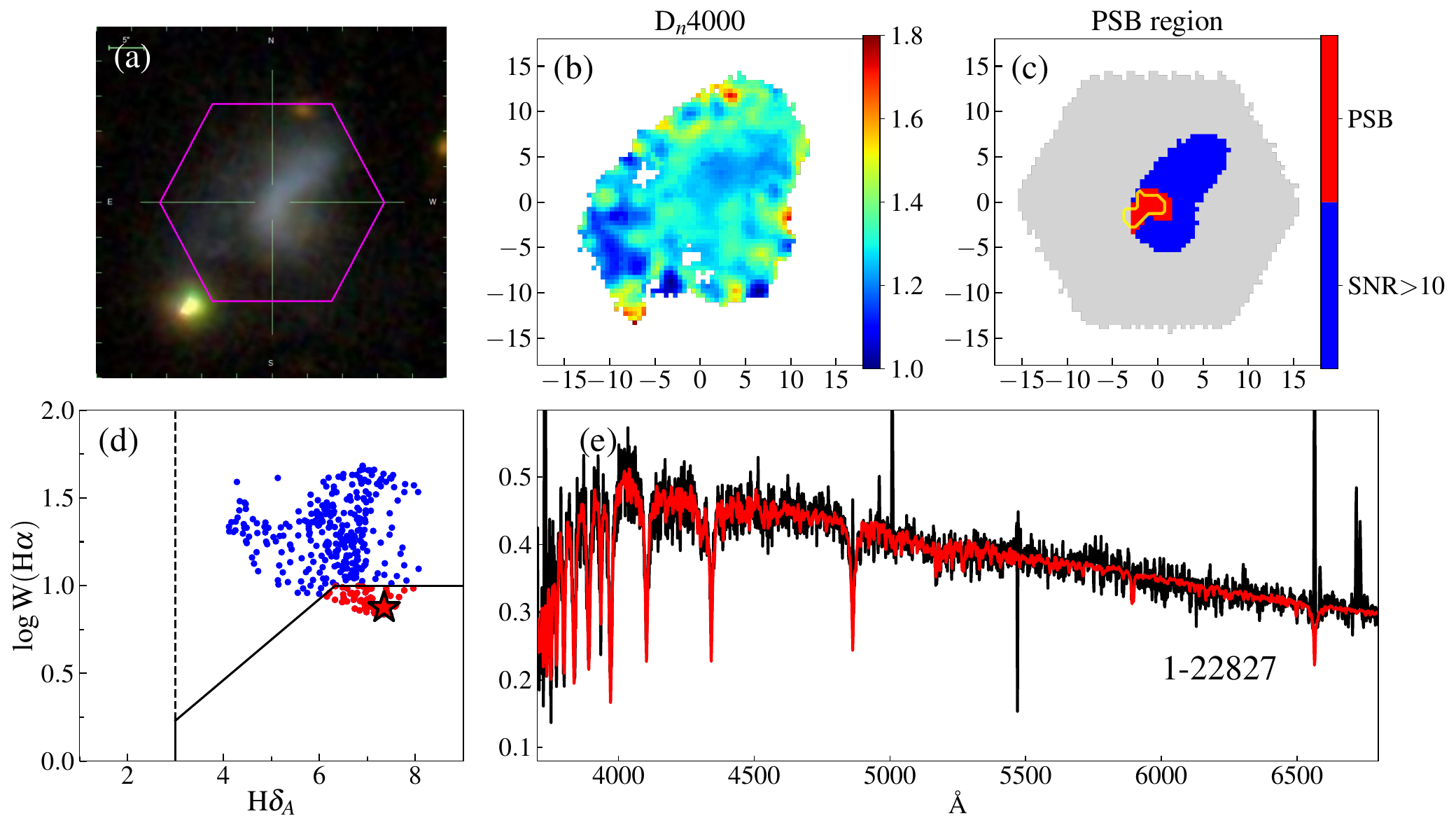}
    \caption{An example of low-mass CPSB galaxy with $\log (M_{*}/\msun) = 9.37$. Panel (a) displays the SDSS $g,r,i$-image. Panel (b) shows the $\dn$ map, where bluer color indicate smaller values and redder color indicate larger values, ranging from 1.0 to 1.8. Panel (c) displays PSB spaxels in red, spaxels of median spectral S/N $>$ 10 per pixel in blue, lower S/N spaxels in gray, and outlines the PSB region defined by \citet{Cheng_2024} in yellow. Panel (d) presents the $\wha$-$\hda$ diagram, where the central spaxel is marked by a red star, PSB spaxels in red, and other spaxels with median spectral S/N per pixel $>$ 10 in blue. Panel (e) shows the central spectrum, with the observed flux in black and the best-fit stellar continuum model in red. MaNGA ID of the galaxy is labeled at the bottom right.}
    \label{fig:7}
\end{figure*}

By visually inspecting the SDSS images of CPSBs that contribute to the peak at the low-mass end and exhibit blue $g-r$ colors, we find they are dominated by the same batch of irregular galaxies. Although early studies have found that the CPSB features are closely associated with merging phenomena \citep{Wild_2009,Snyder_2011,Pawlik_2019,Davis_2019,Chen_2019,Zheng_2020,Ellison_2022}, these low-mass, blue CPSBs appear less consistent with a major merger driven origin, as major mergers would be expected to produce more centrally concentrated, bulge-dominated systems than the irregular galaxies found here. We divide CPSBs into 26 low-mass ($M_{*} < 10^{9.5}\,M_\odot$; hereafter L-CPSB) and 66 high-mass ($M_{*} > 10^{9.5}\,M_\odot$; H-CPSB) subsamples. Fig.~\ref{fig:7} presents an example of a L-CPSB galaxy with $\log (M_{*}/\msun) = 9.37$. Since none of the AGN-PSBs and only three RPSBs are dwarf irregulars, we only compare H-CPSBs with other PSB sub-types in the following analyses to avoid mass-driven biases in subsequent comparisons. The H-CPSBs have a median $\log(M_*/{\rm M_\odot}) \sim 10.12$, which is lower than those of AGN-PSBs and RPSBs by $\sim 0.39$ and $\sim 0.15$ dex, respectively.

\subsection{Kinematics and Morphological Features} \label{sec:kinematics and morphological features}

Previous studies have proposed a variety of mechanisms responsible for triggering the starburst episodes observed in PSBs. Violent processes such as major mergers and strong tidal interactions are among the most efficient drivers, as gravitational torques can funnel large amounts of cold gas into the central regions, igniting intense star formation \citep{Bekki_2005,Wild_2009,Snyder_2011,Pawlik_2019,Davis_2019,Zheng_2020,Ellison_2022,Ellison_2024}. Meanwhile, more moderate pathways, such as acquisition of external gas, can also replenish the cold gas reservoir and induce new cycle of star formation in a less disruptive manner \citep{Kerevs_2005,Chen_2016}. All these processes may not only leave morphological imprints, such as tidal features and disturbed structures, but also manifest kinematic misalignment between stellar and gaseous components \citep{Pawlik_2016,Pawlik_2019,Zhou_2022,Xu_2022}.

To obtain robust measurements of gaseous kinematics, we derive the kinematic position angle (PA) for both stellar (PA$_{\text{star}}$) and gas (PA$_{\text{gas}}$) components in a sample of 7503 emission-line galaxies, using the \texttt{FIT\_KINEMATIC\_PA} Python module \citep{Krajnovic_2006}. The kinematic PA is defined as the counterclockwise angle between celestial north and a line bisecting the velocity field. Stringent spaxel selection criteria were applied to ensure the reliability of velocity field fitting: for the gas component, only spaxels with $\ha$ S/N $>$ 3 were retained; for stellar component, we impose a threshold of median spectral S/N $>$ 3 per pixel. Following the criteria of \cite{Zhou_2022}, we identify 487 misaligned galaxies with $\Delta \mathrm{PA} = |\mathrm{PA_{star}} - \mathrm{PA_{gas}}| > 30^\circ$ and robust PA measurements (i.e., $\mathrm{PA_{err}} \leqslant 20^\circ$).

In addition, we searched for faint signatures of interactions or merger remnants using imaging data from the Dark Energy Spectroscopic Instrument (DESI) Legacy Survey \citep{Dey_2019}. DESI images reach a depth that is typically 1 $\sim$ 2 magnitudes fainter than SDSS in $g$- and $r$-bands. Following the methodology outlined in \cite{Li_2021}, we convolve the $r$-band images with a Gaussian kernel to match the spatial resolution of the $g$-band images, then stacked the two bands to improve the S/N. From these stacked images, we identified four categories of interaction or merger remnant features: (1) isolated galaxies displaying tidal features, (2) morphologically distorted galaxies with nearby companions, (3) galaxies hosting shell-like structures, and (4) galaxies with extended asymmetric stellar halos. Statistical results for misalignment and interaction features are summarized in Table~\ref{tab:2}, whereas the specific features of individual galaxies are documented in Tables~\ref{tab:cpsb}, \ref{tab:rpsb}, and \ref{tab:agnpsb}.

It is clear that all types of PSBs show higher proportions of both interaction/tidal features and gas--star misalignment than their controls, indicating a connection between PSB phenomena and external processes. Compared with RPSBs and AGN-PSBs, a larger fraction of L-CPSBs and H-CPSBs lack emission lines. Among the 26 L-CPSBs, 4 galaxies exhibit interaction/tidal features, whereas only 1 galaxy in the control sample shows such signatures. No gas--star kinematic misalignment is detected in either the L-CPSB sample or its control sample. For the 66 H-CPSBs, 24 present interaction/tidal features and 21 show gas--star kinematic misalignment, whereas the corresponding control sample contains only 5 galaxies with interaction/tidal features and 3 with gas–star kinematic misalignment. None of the RPSBs and AGN-PSBs are lineless systems. Among the 89 RPSBs, 26 galaxies exhibit interaction/tidal features and 12 show gas--star kinematic misalignment, compared to 5 and 2 galaxies in their control sample, respectively. For AGN-PSBs, 12 out of 48 galaxies display interaction/tidal features and 6 show gas--star kinematic misalignment, compared to 3 and 6 galaxies in their corresponding control sample.

\begin{table*}[htbp]
\centering
\caption{The Statistics of lineless galaxies (No-EML), galaxies with interaction or tidal features, as well as those exhibiting misalignment phenomena in different PSB sub-types and their controls. Numbers in parentheses are the overlap between lineless galaxies and those displaying interaction or tidal features, while numbers in square brackets are the overlap between galaxies with misalignment phenomena and those displaying interaction or tidal features.}
\small
\label{tab:2}
\begin{tabular}{lcccc}
\toprule
Type & Total & No-EML & Interaction/Tidal features & Misalignment \\
\midrule
AGN-PSB               & 48 & 0  & 12 & 6 [3] \\
AGN-PSB control       & 48 & 0  & 3  & 6 [0] \\
\midrule
L-CPSB         & 26 & 10 (1) & 4  & 0 \\
L-CPSB control & 26 & 4  & 1  & 0 \\
\midrule
H-CPSB        & 66 & 16 (5)  & 24 & 21 [7] \\
H-CPSB control& 66 & 3  & 5  & 3 [1]  \\
\midrule
RPSB                  & 89 & 0  & 26 & 12 [3] \\
RPSB control          & 89 & 3 (1)  & 5  & 2 [0] \\
\bottomrule
\end{tabular}
\end{table*}

\subsection{Properties of Three Types of PSB Galaxies} \label{sec:properties}

In the subsequent sections, we investigate the global (i.e., $\re$, S$\acute{\rm e}$rsic index and $g-r$ color) and spatially resolved properties (i.e. $\dn$, $\hda$, $\wha$, V$_{\text{star}}$/$\sigma_{\text{star}}$, light-weighted and mass-weighted stellar ages) of the AGN-PSB, CPSB and RPSB hosts. We are interested in understanding the evolutionary pathways and the possible connections among different PSB sub-types.

\subsubsection{Global Properties} \label{sec:global}

\begin{figure*}[htbp]
    \centering
    \includegraphics[width=1\textwidth]{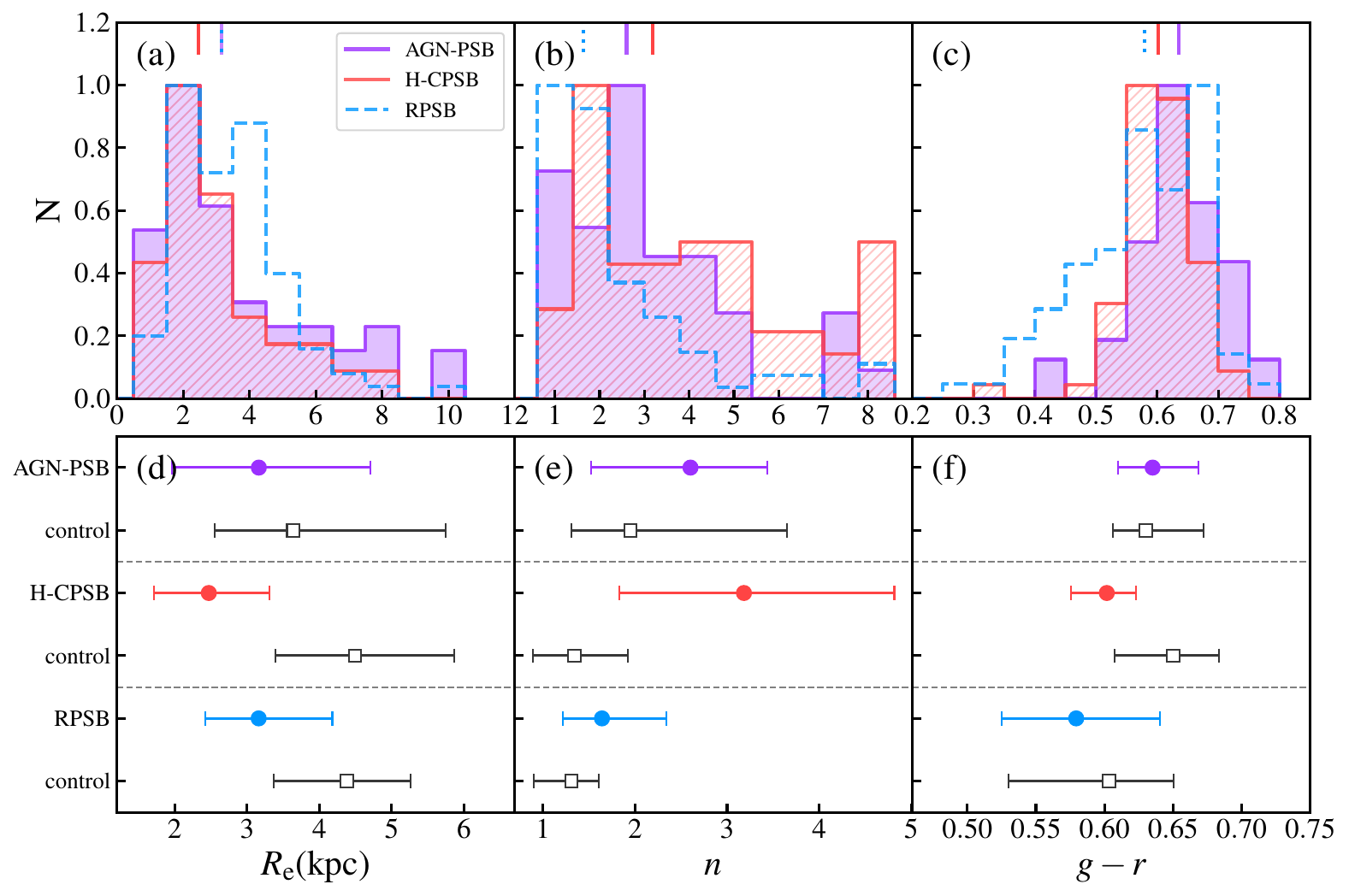}
    \caption{Distributions of global properties for H-CPSBs (red), RPSBs (blue), and AGN-PSBs (purple). The top row shows histograms of $\re$ (panel a), S$\acute{\rm e}$rsic index (panel b), and $g-r$ (panel c). Vertical lines at the top of each panel mark the median of each distribution. The bottom row compares each PSB sub-type with its control sample in $\re$ (panel d), S$\acute{\rm e}$rsic index (panel e), and $g-r$ color (panel f). Colored solid circles and black open squares denote the medians of the PSB sub-types and their controls, respectively. Error bars indicate the 30th to 70th percentiles of the distributions.}
    \label{fig:8}
\end{figure*}

The top row of Fig.~\ref{fig:8} presents the distributions of global properties for AGN-PSBs (purple), H-CPSBs (red), and RPSBs (blue), including $\re$ (panel a), S$\acute{\rm e}$rsic index $n$ (panel b), and $g-r$ (panel c). Vertical lines at the top of each panel mark the median value of each distribution. The median $\re$ of both AGN-PSBs and RPSBs is $\sim$3.2~kpc, which is 0.7~kpc larger than H-CPSBs. Panel (b) demonstrates that H-CPSBs have the highest median S$\acute{\rm e}$rsic index ($n \sim 3.2$), indicating concentrated light profiles and bulge-dominated morphologies. By contrast, RPSBs display disk-dominated morphologies with a median S$\acute{\rm e}$rsic index of $n \sim 1.6$. The median S$\acute{\rm e}$rsic index of AGN-PSBs falls between these two populations with a median value of 2.6. Finally, panel (c) shows that the median $g-r$ colors of the three subtypes differ only slightly, clustering around $\sim$0.6. However, RPSBs exhibit the broadest and bluest color distribution, which can be attributed to their relatively ongoing star formation in the central regions. In contrast, AGN-PSBs are the reddest, while H-CPSBs lie between the two populations. 

Building on the distributions presented above, we performed a Kolmogorov--Smirnov (K--S) test to statistically verify the differences between the samples, and we adopt a threshold of $p=0.05$. When comparing AGN-PSBs and H-CPSBs, the K--S test yields $p=0.191$, $p=0.236$ and $p=0.005$ for the distributions of $\re$, S$\acute{\rm e}$rsic index, and $g-r$ color, respectively. Meanwhile, for the comparisons between AGN-PSBs and RPSBs, the corresponding $p$-values are 0.214, 0.015 and 0.003 for these three parameters. These results indicate that, with the data at hand, the $\re$ distributions of AGN-PSBs are not statistically distinguishable from those of either H-CPSBs or RPSBs. For the S$\acute{\rm e}$rsic index, AGN-PSBs show a statistically significant difference from RPSBs, but not from H-CPSBs. In contrast, their $g-r$ colors differ significantly from both H-CPSBs and RPSBs. We caution against overinterpreting the presented statistical results. Direct visual inspection of Fig.~\ref{fig:8}(a) suggests that AGN-PSBs follow a distinct distribution relative to RPSBs, yet the K--S test returns no statistically significant difference in $\re$. This apparent inconsistency likely stems in part from our limited sample size: resolving marginal statistical offsets reliably demands a substantially larger dataset. Accordingly, a non-significant K--S outcome cannot be taken as definitive evidence for intrinsic equivalence between the two parent distributions; it merely indicates that the available data lack sufficient statistical power to discriminate between them.

In the bottom row of Fig.~\ref{fig:8}, we compare $\re$, S$\acute{\rm e}$rsic index $n$ and $g-r$ color between each PSB sub-type and their controls. Colored solid circles and black open squares denote the medians of the PSB sub-types and their controls, respectively, while error bars indicate the 30th to 70th percentiles of the distributions. The median value of $\re$ is 2.5~kpc for the H-CPSBs and 4.5~kpc for the relevant controls. For the RPSBs, the median $\re$ is 3.2~kpc, while it is 4.4~kpc for the control galaxies. The median $\re$ is 3.2~kpc for AGN-PSBs and 3.6~kpc for their controls. It is clear that the difference in $\re$ between each PSB sub-type and its relevant control decreases from 2.0~kpc for H-CPSBs to 1.2~kpc for RPSBs, and it is only 0.4~kpc for AGN-PSBs. A similar trend is found in S$\acute{\rm e}$rsic index. $n$ = 2 is the often used proxy for bulge versus disk dominated galaxies. All the PSB sub-types have higher $n$ than their control galaxies. Both H-CPSBs and AGN-PSBs are more spheroid dominated, while their control galaxies seem to be disk like. However, the difference in $n$ is 1.8 for H-CPSBs, which is much larger than that of AGN-PSBs. Both RPSBs (median $n$ = 1.6) and their controls (median $n$ = 1.3) are disk dominated with $n < 2$. In terms of the $g-r$, only H-CPSBs are significantly bluer than their control galaxies, with a median offset of 0.07 mag. RPSBs show only a mild color difference of 0.02 mag relative to their controls, while the $g-r$ colors of AGN-PSBs and their control galaxies are nearly indistinguishable.

Taken together, these trends suggest that the luminosity enhancement in H-CPSBs is centrally concentrated towards the bulge, whereas in RPSBs and AGN-PSBs the star formation is not centrally concentrated but instead extends over a broader radial range. To further test this scenario, we examine spatially resolved galaxy properties in the following sections, focusing on radial gradients of key parameters to clarify the physical differences between the central and outer regions.

\subsubsection[Radial Gradients of Dn, HdA, WHa]%
{Radial Gradients of $\dn$, $\hda$ and $\wha$} \label{sec:radial1}

\begin{figure*}[htbp]
    \centering
    \includegraphics[width=1\textwidth]{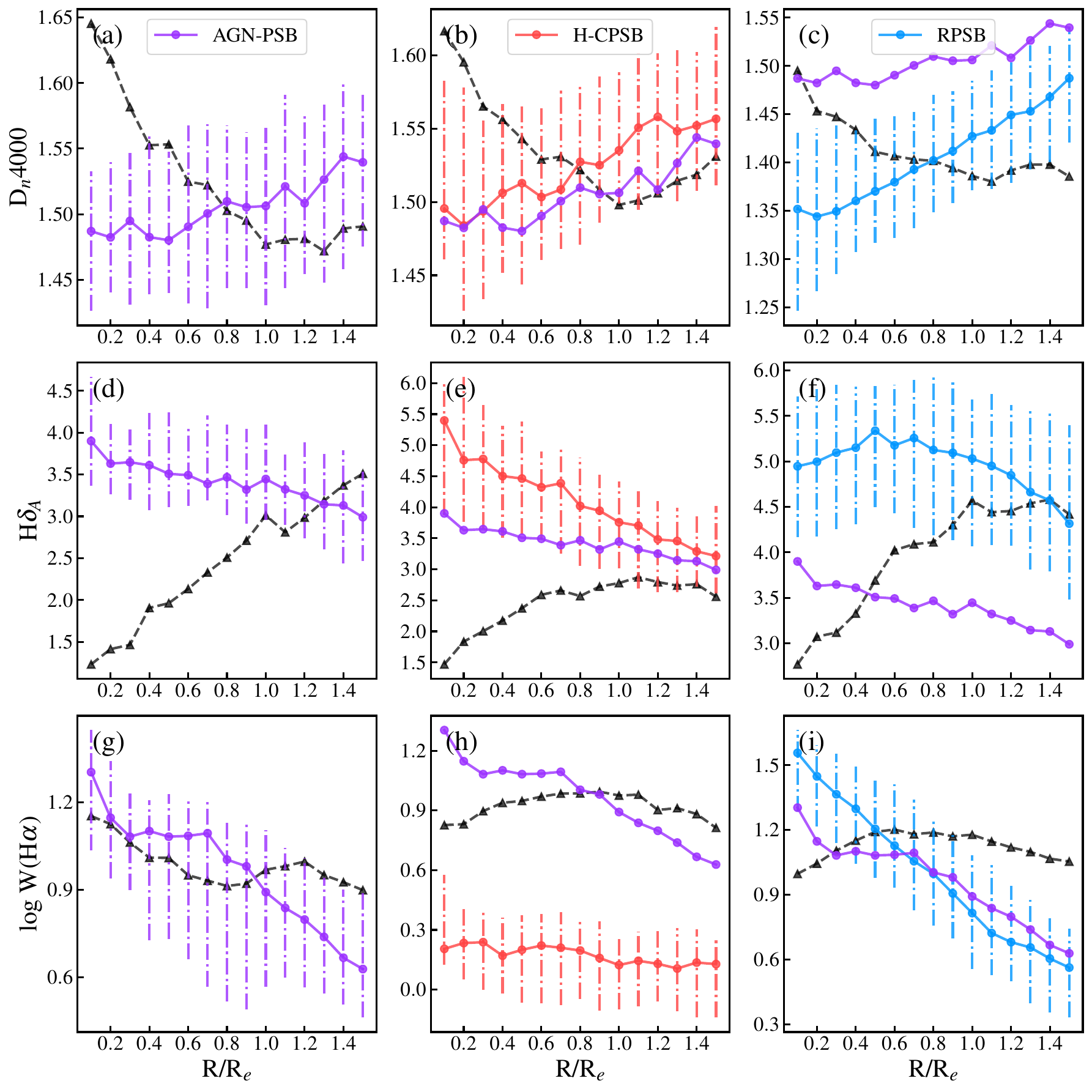}
    \caption{The median radial profiles of $\dn$ (top row), $\hda$ (middle row), and $\wha$ (bottom row) as a function of radius for AGN-PSB (left column; purple solid line), H-CPSB (middle column; red solid line), RPSB (right column; blue solid line) and their control samples (black dashed line). The error bars show the 30th to 70th percentile of the distribution for PSB sub-types. For direct comparison, the median gradient of AGN-PSBs is overlaid in the middle and right columns.}
    \label{fig:9}
\end{figure*}

Fig.~\ref{fig:9} presents the median radial profiles of $\dn$ (top row), $\hda$ (middle row), and $\wha$ (bottom row) for AGN-PSBs (left column; purple solid line), H-CPSB (middle column; red solid line), RPSB (right column; blue solid line), and their respective control samples (black dashed line). Error bars denote the 30th to 70th percentile range of the PSB distributions. For direct comparison, the median gradient of AGN-PSBs is overlaid in the middle and right columns.

As found in panel (a), AGN-PSBs exhibit positive radial gradients in $\dn$, implying younger stellar populations in the central regions and older populations in their outskirts. By contrast, the control galaxies display distinctly negative $\dn$ gradients. An AGN-PSB is essentially a CPSB with concurrent central black hole activity, and all of our AGN-PSB hosts are Type 2 Seyfert/LI(N)ER galaxies, in which the AGN continuum contribution is extremely small, rarely exceeding 5\% \citep{Schmitt_1999,Kauffmann_2003}. Thus the influence of the AGN continuum on the $\dn$ measurement is negligible. The pronounced negative $\dn$ gradient observed in the AGN-PSB control sample further supports this interpretation. We note here that $\oiii$ luminosity was adopted as a tracer of central black hole activity when constructing the AGN-PSB control sample. The obvious difference in $\dn$ radial profiles between AGN-PSBs and their controls provides additional evidence that the host galaxy properties of AGN-PSBs are distinct from those of AGN hosts without PSB features. In panel (b) and (c), the $\dn$ profiles of H-CPSBs, RPSBs and their controls are consistant with \citet{Chen_2019}, we suggest readers to read their section 3.2 for detailed explanations. We overplot the median distributions of AGN-PSBs as purple lines in both two panels for comparison. It is clear that AGN-PSBs and H-CPSBs exhibit nearly identical $\dn$ values throughout their entire radial extent. Whereas these values are systematically higher than those of RPSBs, suggesting that RPSBs represent the youngest systems in statistic.

In panel (d), AGN-PSBs exhibit negative $\hda$ gradients, in contrast to the positive gradients shown by the controls. Again, panel (e) and (f) agree with \citet{Chen_2019}. Although both AGN-PSBs and H-CPSBs show negative $\hda$ gradients, the $\hd$ absorption strength in the central region of AGN-PSBs is substantially smaller than that of H-CPSBs. This discrepancy diminishes gradually with the increasing radial distance in the outer regions and disappears at the outermost radius. The potential explanations for the weaker central $\hda$ absorption in AGN-PSBs are proposed as follows: (1) it is intrinsic, indicating that the recent past starburst activity of AGN-PSB was inherently weaker than that of H-CPSBs; (2) it is affected by AGN emission lines, which suppress or obscure the $\hd$ absorption signal. In contrast, RPSBs show roughly flat and overall stronger $\hda$ profiles than AGN-PSBs.

Panel (g) shows that AGN-PSBs exhibit strongly negative $\wha$ gradients. Their values are comparable to those of controls in the inner regions but become significantly lower in the outskirts. The similarity in the center may be attributed to considerable AGN activity present in both AGN-PSBs and their controls, while the outer-region discrepancy likely arises from ongoing star formation in the controls. This interpretation is supported by panel (a), which shows that the control sample has a smaller $\dn$ than AGN-PSBs in the outskirts, indicating younger stellar populations. As expected, the profiles of H-CPSBs and RPSBs are in agreement with \citet{Chen_2019}. In panel (h), H-CPSBs show significantly lower $\wha$ from the inner to outer regions than AGN-PSBs, indicating strongly suppressed ongoing star formation activity. In panel (i), AGN-PSBs and RPSBs seem share the same $\wha$ profiles. Using the widely adopted BPT diagram based on $\siil/\ha$ versus $\oiii/\hb$ flux ratios \citep{Kewley_2006}, we find that the central $\ha$ emission in RPSBs is predominantly powered by star formation: 79 of the 89 RPSBs are classified as SF, while 3 are classified as Seyfert and 7 as LINER. By contrast, the central $\ha$ emission in AGN-PSBs arises from central black hole activity. Thus the strong Balmer absorption in the central regions of the AGN-PSBs constitutes evidence of rapid quenching of star formation.

\subsubsection{Radial Gradients of Mass- and Light-Weighted Ages} \label{sec:radial2}

\begin{figure*}[htbp]
    \centering
    \includegraphics[width=1\textwidth]{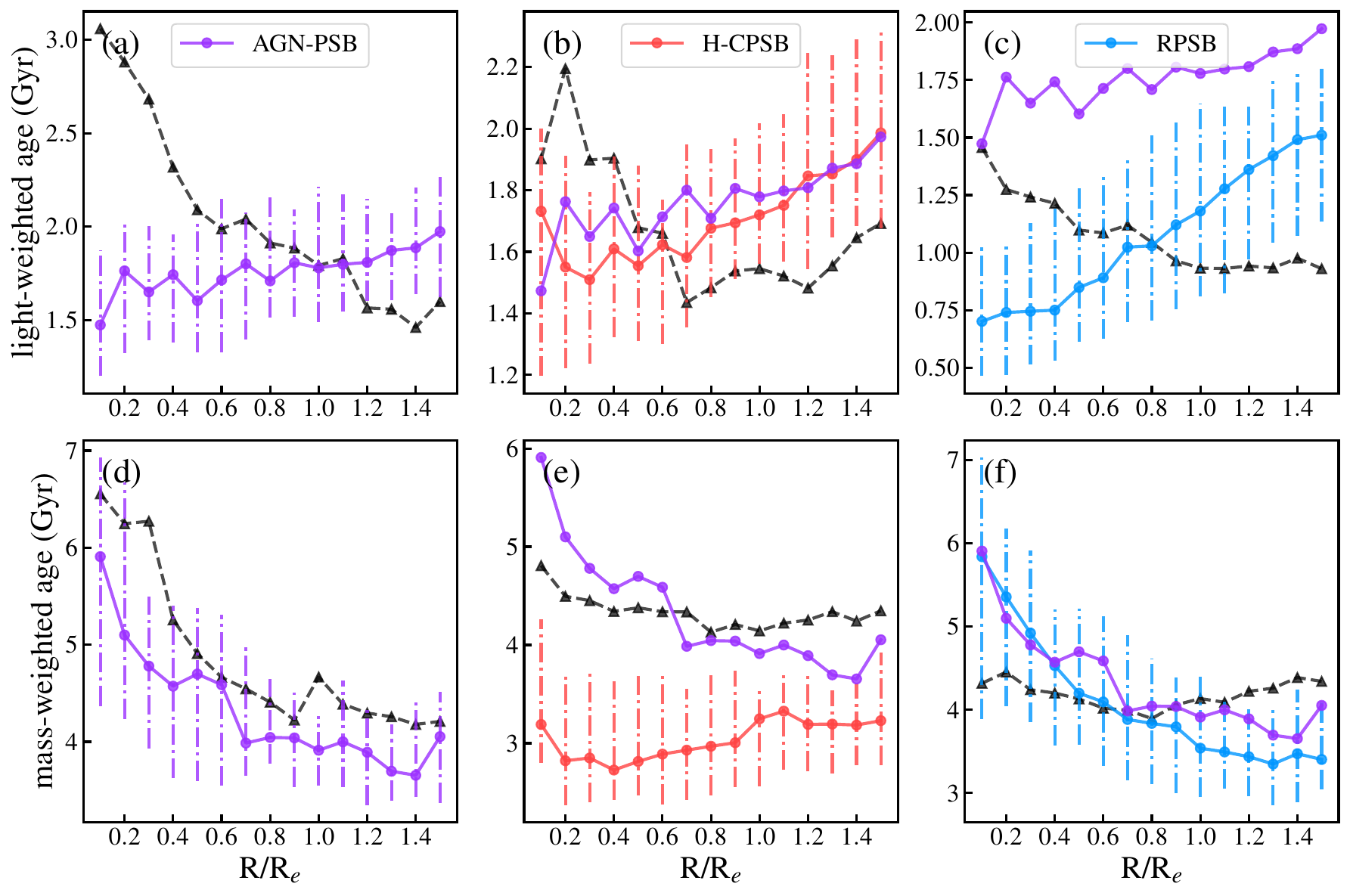}
    \caption{The median radial profiles of light-weighted age (top row) and mass-weighted age (bottom row) as a function of radius for AGN-PSB (left column; purple solid line), H-CPSB (middle column; red solid line), RPSB (right column; blue solid line) and their control samples (black dashed line). The error bars show the 30th to 70th percentile of the distribution for PSB sub-types. For direct comparison, the median gradient of AGN-PSBs is overlaid in the middle and right columns.}
    \label{fig:10}
\end{figure*}

Fig.~\ref{fig:10} presents the median radial profiles of light-weighted stellar age (top row) and mass-weighted stellar age (bottom row). The symbols and lines are color-coded as the same in Fig.\ref{fig:9}.

Since $\dn$ serves as a robust indicator of light-weighted stellar age, the spatial distribution of light-weighted age in Fig.~\ref{fig:10} is highly consistent with that of $\dn$ presented in the top row of Fig.~\ref{fig:9}. In contrast to light-weighted age, which is sensitive to recent star formation, mass-weighted age is dominated by the accumulated mass of long-lived low- to intermediate-mass stars. Panel (d) shows that AGN-PSBs exhibit negative mass-weighted age gradients. And the values of mass-weighted age show only a mild decrease relative to their controls. Panels (e) and (f) present results in agreement with \citet{Chen_2019}, showing that H-CPSBs have totally distinct profiles from RPSBs. When we bring in the new AGN-PSB category, we find somewhat surprisingly that they closely resemble RPSBs: both are characterized by an old ($\sim$6 Gyr) stellar population in the center that decreases with increasing radius. This suggests that the observed recent star formation constitutes merely a "frosting" atop a predominantly old stellar population. In contrast, the prominent younger ages of H-CPSBs at all radii reflect a distinctly different evolutionary history, with a higher fraction of stellar mass formed more recently throughout the entire galaxy. 

\subsubsection[Radial Gradients of V/sigma] 
{Radial Gradients of $V_{\text{star}}/\sigma_{\text{star}}$} \label{sec:radial3}

The ratio of ordered rotational to random dispersive motion of stars in galaxies, ($V_{\text{star}}/\sigma_{\text{star}}$), is a key diagnostic for galaxy dynamical states. This kinematic parameter correlates strongly with fundamental properties (e.g., luminosity, stellar mass; \citealt{illingworth_1977, Davies_1983, Emsellem_2011, Brough_2017, van_de_Sande_2017, Veale_2017, Green_2018}), implying an intrinsic link between stellar mass assembly and angular momentum evolution across cosmic time. Violent events like major mergers disrupt rotation, stir stellar motions, and increase velocity dispersion, thereby lowering $V_{\text{star}}/\sigma_{\text{star}}$ and forming bulge-dominated, dynamically hot galaxies. Minor mergers, though less disruptive, asymmetrically alter mass and angular momentum distributions. However, mergers are by no means the sole physical process at work; continuous gas accretion and star formation can also alter the morphological and kinematic properties of galaxies \citep{Naab_2014}.

\begin{figure*}[htbp]
    \centering
    \includegraphics[width=1\textwidth]{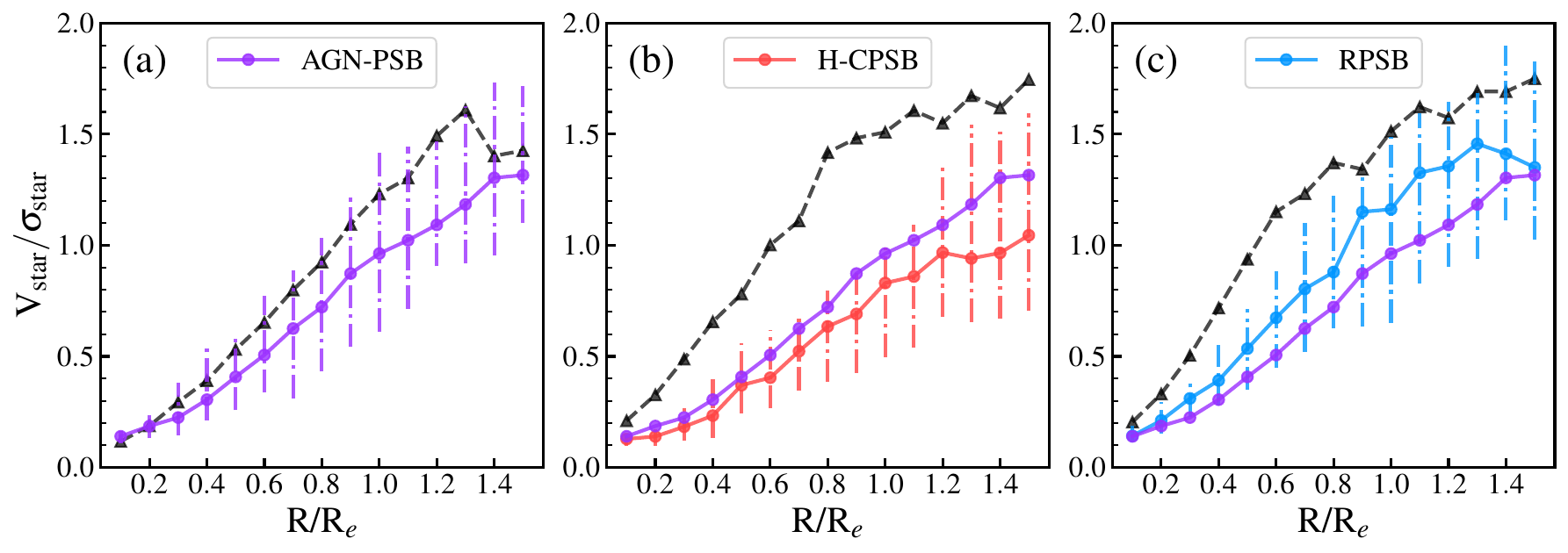}
    \caption{The median radial profiles of V$_{\text{star}}$/$\sigma_{\text{star}}$ as a function of radius for AGN-PSB (left column; purple solid line), H-CPSB (middle column; red solid line), RPSB (right column; blue solid line) and their control samples (black dashed line). The error bars show the 30th to 70th percentile of the distribution for PSB sub-types. For direct comparison, the median gradient of AGN-PSBs is overlaid in the middle and right columns.}
    \label{fig:11}
\end{figure*}

Fig.~\ref{fig:11} presents the median radial profiles of $V_{\text{star}}/\sigma_{\text{star}}$. A higher (lower) $V_{\text{star}}/\sigma_{\text{star}}$ ratio denotes stronger (weaker) rotational support, enabling us to probe the potential differences in the formation and interaction histories across PSB sub-types. A comparison with fig.~2 of \citet{Emsellem_2007} shows that the radial $V_{\rm star}/\sigma_{\rm star}$ profiles of our samples are characteristic of fast rotators. This result is within our expectation, since slow rotators are typically more massive and relatively rare in the local universe. All three PSB sub-types have lower $V_{\text{star}}/\sigma_{\text{star}}$ ratios compared to their respective controls, with the largest difference detected specifically between H-CPSBs and their controls. Among the PSB sub-types, H-CPSBs exhibit the smallest $V_{\text{star}}/\sigma_{\text{star}}$, followed by AGN-PSBs, while RPSBs have the largest ratios. The lowest $V_{\text{star}}/\sigma_{\text{star}}$ ratios of H-CPSBs among all PSB sub-types and control samples suggest that these galaxies have undergone more frequent and/or violent mergers, interactions or gas accretion processes \citep{Lagos_2018}. This is entirely consistent with the statistics of external processes reported in Tab.~\ref{tab:2}, which confirm the highest interaction/merger and gas--star misalignment fractions for H-CPSBs.

Moreover, such external processes can drive gas inflows---a key mechanism for interpreting the observed positive $\dn$ gradients in all PSB sub-types. These inflows induce gas accumulation in the central regions, which in turn triggers intense central star formation and produces the young stellar populations (low $\dn$) detected in the central regions. For instance, the infall of counter-rotating gas---originating either from gas-rich dwarf companions or from the cosmic web---can redistribute angular momentum through collisions between the newly accreted gas and the pre-existing gaseous component. This interaction can significantly enhance radial gas inflow toward the central regions \citep{Chen_2016}. Other mechanisms such as galactic bars may also play a role \citep{Hawarden_1986,Lin_2017,Chown_2019}.

\section{Discussion} \label{sec:discussion}

PSBs represent a transient phase in galaxy evolution, characterized by the rapid shutdown of star formation following a recent starburst. Understanding the origin of PSBs therefore requires addressing two closely related questions: what physical processes trigger the starburst, and what mechanisms subsequently quench star formation on short timescales. Actually, both questions can be reframed in terms of the cycle of cold gas. The onset of the starburst demands an efficient supply or redistribution of cold gas, while rapid quenching requires this gas to be consumed, expelled, or heated to prevent further star formation. Identifying how these processes operate in PSBs is essential for constraining their evolutionary pathways and contributing to the theories of galaxy evolution.

\subsection{Morphology Evolution} \label{sec:morph evo}

A number of previous studies have shown that PSBs are more compact and bulge-dominated than coeval star-forming or quiescent galaxies \citep{Whitaker_2012,Wu_2020,Chen_2022,Setton_2022}. However, the PSB selection is based on single-fiber spectroscopic observations, and therefore primarily reflect the properties of CPSBs. 

Fig.~\ref{fig:8} compares $\re$ and S$\acute{\rm e}$rsic index $n$ between PSB sub-types and their respective controls. The significant differences in both $\re$ and S$\acute{\rm e}$rsic index $n$ between H-CPSBs and their controls are consistent with a picture in which major mergers or other gravitational interactions efficiently remove gas angular momentum and drive radial gas inflows, thereby triggering centrally concentrated starbursts and substantial morphological transformation in the majority of the sample. In contrast, the milder gas accretion and bar-driven gas inflow mechanisms operating in RPSBs and AGN-PSBs only modestly perturb their morphologies, resulting in far smaller differences between these two sub-types and their corresponding control samples. This interpretation is supported by the interaction statistics presented in Tab.~\ref{tab:2}. Considering the total interaction fraction (either merger/tidal features or gas--star kinematic misalignment), the corresponding values are $\sim$58\%, $\sim$39\%, and $\sim$31\% for H-CPSBs, RPSBs, and AGN-PSBs, respectively. We quantify these difference using Fisher’s exact test for binomial proportions, adopting $p < 0.05$ as a threshold for statistical significance. The results show that H-CPSBs exhibit a substantially higher overall interaction fraction compared with both RPSBs ($p=0.034$) and AGN-PSBs ($p=0.008$). By contrast, no statistically meaningful divergence emerges between RPSBs and AGN-PSBs ($p=0.457$).

\citet{Cheng_2024} investigated the stellar mass--size relation of CPSBs, RPSBs and IPSBs selected from the final data release of the MaNGA survey (see their fig~3). They found that only CPSBs at intermediate stellar masses ($9.5 < \log(M_{*}/\msun) < 10.5$) exhibit systematically smaller sizes than their control galaxies, with a typical offset of $\Delta\log\re \sim 0.2$, whereas CPSBs at lower mass ($\log(M_{*}/\msun) < 9.5$) or higher mass ($\log(M_{*}/\msun) > 10.5$) closely follow the mass--size relations of their controls. H-CPSBs in our sample have a median stellar mass of $\log(M_{*}/\msun) \approx 10.1$ and a median effective radius of $\log \re \sim 0.39$~dex, which is approximately 0.26~dex smaller than that of their controls, in agreement with the results of \citet{Cheng_2024}. The lack of difference in $\re$ among CPSBs with $\log(M_{*}/\msun)<9.5$, which we define as L-CPSBs, may indicate that their evolution is driven by processes acting on more local scales like feedback from stellar winds and supernovae, rather than global mechanisms that can strongly reshape the overall galaxy structure.

\subsection{Evolutionary Pathways} \label{sec:evo path}

In this section, we focus on the potential evolutionary connections among the three PSB sub-types. As shown in Fig.~\ref{fig:5}, 77\% CPSBs lie in the green valley, implying that they represent a transitional phase between star-forming and quiescent galaxies. The evolutionary pathways leading to the PSB phase, however, are likely far more complex. Based on SDSS single-fiber spectroscopy, \citet{Pawlik_2018} proposed three distinct evolutionary pathways for CPSBs: (1) gas-rich major mergers driving blue-cloud galaxies onto the red sequence; (2) less intense events inducing cyclic evolution in low-mass star-forming  galaxies($9.5<\log(M_{*}/\msun)<10.5$) within the blue cloud; and (3) similarly mild processes triggering cyclic evolution in massive quiescent galaxies ($\log(M_{*}/\msun)>10.5$) within the red sequence. Galaxies following each of these channels may pass through a PSB phase. Notably, galaxies undergoing the first pathway account for approximately 60–70\% of all the CPSBs in their sample, a fraction comparable to the $\sim$ 63\% merger/interaction and gas--star misalignment fraction observed in our H-CPSB sample (Tab.~\ref{tab:2}).

With the development of spatially resolved IFU surveys, PSB signatures can now be detected beyond the central regions of galaxies \citep{Chen_2019}. Regarding RPSBs, previous studies have drawn inconsistent conclusions about their origins and evolutionary pathways. \citet{Chen_2019} reported discrepant radial profiles in the mass-weighted stellar age and $V_{\rm star}/\sigma_{\rm star}$ between CPSBs and RPSBs, suggesting distinct evolutionary channels for the two populations. In contrast, \citet{Cheng_2024} found that both CPSBs and RPSBs undergo an outside-in quenching mode, and thus proposed that RPSBs may evolve into CPSBs as the quenching process proceeds inward. \citet{Chen_2019} investigated the distribution of galaxy spaxels in the $\hda$--$\wha$ plane and identified two RPSB categories (see their fig~8): Type~I RPSBs appear to undergo an outside-in quenching, with comparable quenching timescale both in center and outer; Type~II RPSBs are consistent with a range of quenching timescales, with the outer regions quenching more rapidly than the inner regions. \citet{Leung_2025} added two additional categories of RPSBs (see their fig.~8) based on the work of \citet{Chen_2019}: Type~III features globally synchronized quenching across radii, but the center—preceded by extended continuous star formation—may no longer meet PSB criteria at the present epoch; Type~IV has a long-quenched (or earlier-quenching) center, while only the outskirts undergo a later, gas-fueled rejuvenation episode that quenches after the gas is exhausted. In a sample of 37 RPSBs, \citet{Leung_2025} classified them into 5 Type~I, 18 Type~II, 11 Type~III, and 2 Type~IV systems. Through an analysis of the SFHs of the central SF and outer PSB regions in RPSBs, they suggest that only Type~I RPSBs (13.5\% of the sample) are likely progenitors of CPSBs. In the current work, we reinforce the conclusion of \citet{Chen_2019} using a sample $\sim 3$ times larger. The median mass-weighted stellar ages of RPSBs indicate a substantial old stellar population in their centers ($\sim6$~Gyr), which decreases smoothly with radius to $\sim3.5$~Gyr at $R = 1.5\re$. In contrast, CPSBs exhibit nearly constant mass-weighted ages of $\sim3$~Gyr across the entire galaxy, implying a fundamentally different SFH in which a larger fraction of the stellar mass formed more recently at all radii. These divergent radial gradients in the mass-weighted ages of CPSBs and RPSBs demonstrate that the two populations have vastly different SFHs, making it unlikely for most RPSBs to evolve into CPSBs via secular processes alone, which is totally consistent with the result of \citet{Leung_2025}.

\begin{figure*}[htbp]
    \centering
    \includegraphics[width=1\textwidth]{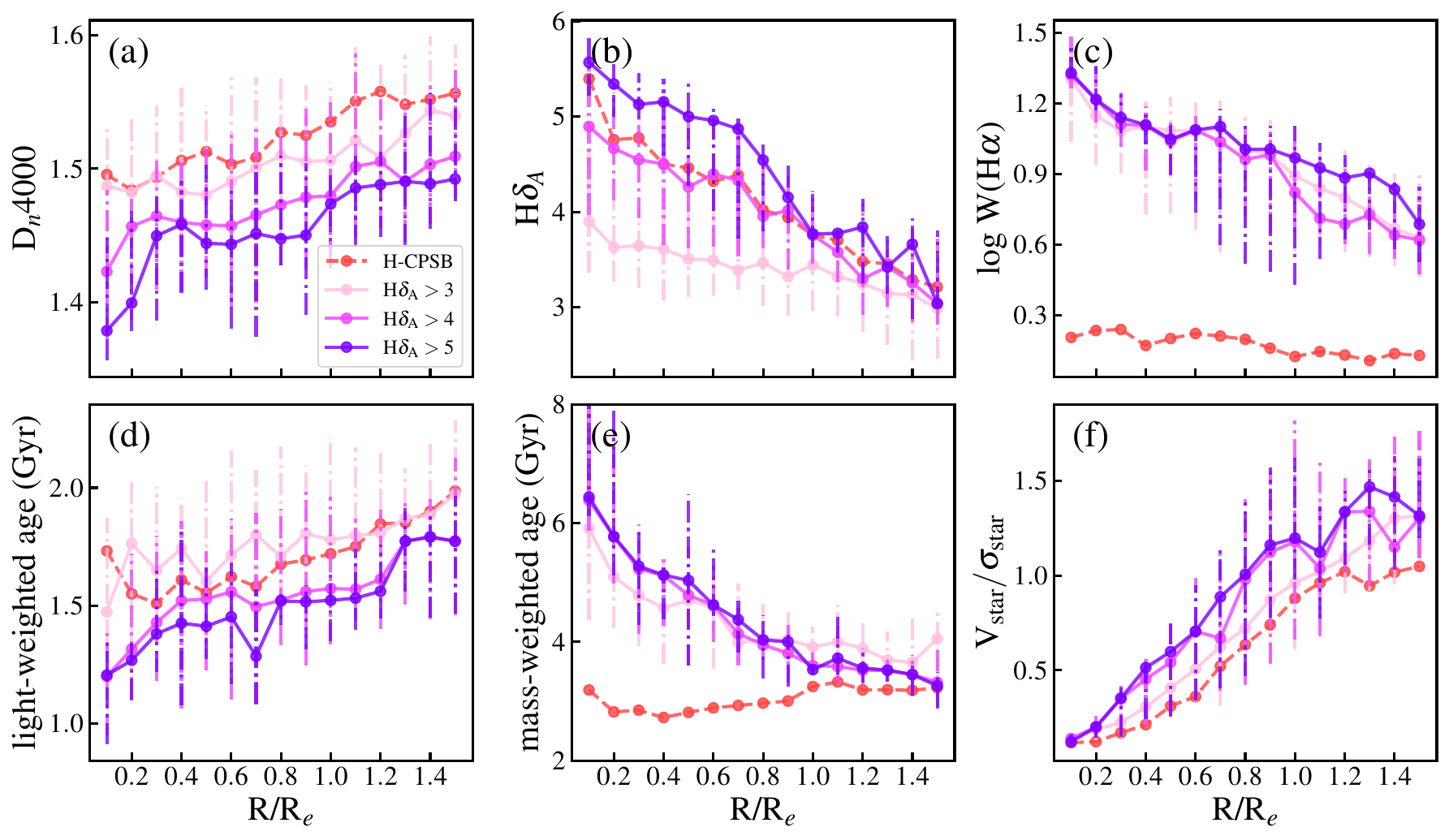}
    \caption{Median radial profiles of $\dn$ (a), $\hda$ (b), $\log\wha$ (c), light-weighted age (d), mass-weighted age (e), and $V_{\mathrm star}/\sigma_{\mathrm star}$ (f) as a function of radius for H-CPSBs (red), AGN-PSBs selected with $\mathrm{H\delta_A}>3$ (pink), 4 (magenta), and 5 (purple), respectively.}
    \label{fig:12}
\end{figure*}

For the new AGN-PSB population proposed in this work, we argue that the vast majority of them cannot evolve into the CPSB population. This is for the same reason as that used to rule out a dominant evolutionary connection between RPSBs and CPSBs—specifically, AGN-PSBs exhibit median mass-weighted stellar ages that are markedly different from those of CPSBs (see Fig.~\ref{fig:10}). Given that our PSB selection relies on $\hda$, we proceed to examine the robustness of this conclusion against variations in the adopted $\hda$ threshold. To this end, we rerun the AGN-PSB selection with three alternative cuts: $\mathrm{H\delta_A}>3$, 4, and 5~$\mathrm \AA$, returning sample sizes of 48, 20, and 11 galaxies respectively. The resulting radial profiles are compared with those of H-CPSBs in Fig.~\ref{fig:12}.

As expected, panel (b) shows that increasing the $\hda$ threshold naturally produces stronger $\hda$ profiles. Panels (a) and (d) further show that stronger-$\hda$ AGN-PSBs have lower $\dn$ values and younger light-weighted stellar ages at all radii. In contrast, panels (c), (e), and (f) show that the radial profiles of $\log\wha$, mass-weighted age, and $V_{\mathrm star}/\sigma_{\mathrm star}$ are not sensitive to the adopted $\mathrm{H\delta_A}$ threshold. In particular, AGN-PSB samples selected with different $\mathrm{H\delta_A}$ thresholds remain clearly distinct from H-CPSBs in mass-weighted age. These findings indicate that the vast majority of AGN-PSBs within our sample are unlikely to evolve into H-CPSBs, irrespective of the adopted $\hda$ cutoff. Importantly, we stress that this remains a statistical inference and cannot exclude the evolutionary pathway whereby a minor subset of AGN-PSBs eventually transitions into CPSBs.

In contrast, the similarity in mass-weighted age gradients between RPSBs and AGN-PSBs implies a shared past SFH for the two populations, thereby supporting the potential for an evolutionary link between them. The median light-weighted age of RPSBs is $\sim$ 0.75 Gyr in the central region, and it increases with radius, reaching nearly $\sim$ 1.5 Gyr at 1.5$\re$. For AGN-PSBs, the median light-weighted age is $\sim$ 1.5 Gyr in the central region and rises to nearly $\sim$ 2 Gyr in the outer regions. Although a light-weighted age difference is observed between these two populations, this discrepancy can be reasonably explained by the evolution of RPSBs into AGN-PSBs. As cold gas in the central regions of RPSBs continues to flow inward and trigger nuclear activity, central star formation is subsequently suppressed by supernova feedback, AGN feedback, or gas depletion. In the absence of ongoing star formation, the light-weighted ages, which are dominated by young stellar populations, increase rapidly, allowing the stellar populations of RPSBs to evolve toward those observed in AGN-PSBs.

Furthermore, RPSBs and AGN-PSBs exhibit comparable fractions of merger remnant features ($\sim$29\% and $\sim$25\%, respectively) and gas–star kinematic misalignments (both at $\sim$13\%), implying that they likely share similar driving mechanisms for PSB formation. The slightly higher $V_{\rm star}/\sigma_{\rm star}$ of RPSBs compared to AGN-PSBs suggests that RPSBs are more rotation-supported than AGN-PSBs, which can also be naturally interpreted by a possible evolutionary transition from RPSBs to AGN-PSBs. Overall, these results support an evolutionary pathway from RPSBs to AGN-PSBs, which can be further tested in future work through a detailed analysis of the radial dependence of their SFHs.

\subsection{The Role of AGN feedback} \label{sec:agn feedback}

While galaxy mergers and interactions are well-established mechanisms for producing PSBs \citep{Bekki_2005,Wild_2009,Snyder_2011,Pawlik_2019,Davis_2019,Zheng_2020,Ellison_2022,Ellison_2024}, the role of AGN feedback in driving the rapid quenching of star formation remains debated. Numerous studies have reported evidence in favor of AGN-driven quenching through various feedback processes, including radiative heating and kinetic outflows \citep{Goto_2006,Tremonti_2007,Melnick_2015,Baron_2017,Baron_2018}. In contrast, other works argue that AGN outflows are insufficient to globally deplete or heat the cold gas reservoirs \citep{Yesuf_2017,Yesuf_2020,Ellison_2022}. An alternative interpretation is that the processes triggering the starburst, particularly galaxy mergers, also fuel nuclear activity, without AGN feedback being the primary driver of quenching \citep{Hopkins_2006,Pawlik_2018}. This view is supported by the observed $\sim200-300$ Myr delay between the starburst and the onset of AGN activity in the local Universe \citep{Wild_2010,Yesuf_2014}, suggesting that AGN activity in PSBs may be a by-product of gas inflow \citep{Maltby_2019,Lanz_2022}.

According to the $\sii$-based BPT diagrams, 81 of the 89 RPSBs in our sample show that ongoing central star formation persists in the central regions. Given that star formation in the outer regions has recently been quenched but persists in the centers, we argue that AGN feedback — which acts on the central regions first — is not a necessary condition for suppressing star formation and generating PSB features. Furthermore, the large number of IPSBs in our sample indicates that localized PSB regions are a ubiquitous phenomenon, most likely originating from intermediate-age cluster complexes or individual bright star clusters. The formation of PSB regions in IPSBs is thus more likely driven by stellar feedback itself and is unassociated with black hole activity.

For the three sub-types, including H-CPSBs, RPSBs and AGN-PSBs, they all exhibit the characteristic that their central stellar populations are younger than the outskirts. This outside-in growth mode implies the existence of physical mechanisms that drive gas inflow, thereby triggering central star formation. Both external processes (e.g., mergers, interactions, gas accretion) and internal secular evolution (torque induced by non-axisymmetric galactic structures like bars) can drive the redistribution of angular momentum and transport gas inward. Based on a sample of 487 gas--star misaligned galaxies in which the misaligned gas component is inferred to originate primarily from external gas accretion, \citet{Zhou_2025} show that the interaction between accreted and pre-existing gas acts as an angular momentum loss mechanism; as a result, the ensuing gas inflow first triggers star formation in the central region, while a fraction of the gas accretes onto the galactic center and drives subsequent nuclear black hole activities. Thus, a natural link between PSB features and nuclear black hole activities arises from their shared gas reservoir.

However, we cannot exclude the possibility that AGN feedback may still play a role at later stages, potentially contributing to the transition from the PSB phase toward long-term quiescence \citep{Almaini_2025}. Based on the analysis of SFHs, \citet{Leung_2025} showed that in RPSBs not only the outer regions are quenching, but the central regions have also experienced substantial quenching over the past $\sim$2~Gyr, with the mean central sSFR declining by approximately 1~dex during this period. Nevertheless, the central regions of RPSBs are generally not classified as PSB regions because residual star formation remains ongoing. This indicates that the onset of central quenching in RPSBs does not require prior black hole activity. However, as RPSBs evolve into AGN-PSBs, AGN activity may subsequently act to suppress the remaining residual central star formation, allowing galaxies to fully transition into a quiescent state.

\section{Summary} \label{sec:summary}

Based on the traditional PSB selection by \citet{Chen_2019} and a newly defined criterion identifying concurrent PSB and AGN features, we construct the largest spatially resolved sample of PSBs to date using the final data release of the MaNGA survey. Our final sample consists of 48 AGN-PSBs, 92 CPSBs, 89 RPSBs, and 828 IPSBs. We find the global and spatially resolved properties of CPSBs and RPSBs are consistent with the results of \citet{Chen_2019}. In this work, we focus on the properties of AGN-PSBs, comparing them with CPSBs, RPSBs, and control galaxies. There are several important results that can be summarized as:

\begin{enumerate}

\item On the global $\log(M_{*}/\msun)$--$\dn$ diagrams, AGN-PSBs and CPSBs predominantly reside in the green valley, whereas RPSBs predominantly reside on the star-forming main sequence.

\item A subset of CPSBs are low-mass dwarf galaxies, whose formation is unlikely to be major-merger-driven and instead points to localized physical processes. This motivates a division of CPSBs into low-mass CPSBs (L-CPSBs; $\log(M_{*}/\msun)<9.5$) and high-mass CPSBs (H-CPSBs; $\log(M_{*}/\msun)>9.5$).

\item All PSB sub-types exhibit enhanced interaction signatures relative to their control samples, indicating that external processes play an important role in PSB formation. AGN-PSBs have similar merger features/gas--star kinematic misalignment with RPSBs, likely associated with less violent mechanisms (e.g., external gas acquisition), rather than merger-dominated origin.

\item AGN-PSBs show minor differences relative to their control galaxies in structural parameters (e.g., $\re$ and S$\acute{\rm e}$rsic indices). In contrast, the H-CPSBs undergo the most pronounced structural transformation, characterized by smaller $\re$ and higher S$\acute{\rm e}$rsic indices relative to their controls, consistent with merger-driven gas inflows that trigger compact central starbursts and bulge growth.

\item AGN-PSBs exhibit positive $\dn$ gradients compared to their controls, indicating a younger stellar population in the central regions than in the outskirts. Their smaller central $\hda$ values compared to H-CPSBs suggest that the recent starburst/quenching episode in AGN-PSBs was less intense and resulted in weaker PSB signatures. In contrast, the strong central $\ha$ emission is likely dominated by AGN-powered emission lines.

\item The similarity in the radial profiles of mass-weighted age and $V_{\rm star}/\sigma_{\rm star}$ between AGN-PSBs and RPSBs, together with the systematically younger ages of RPSBs, suggests an evolutionary sequence in which RPSBs represent an earlier stage that can evolve into AGN-PSBs following continued gas inflow and the onset of nuclear activity. In contrast, the majority of H-CPSBs likely follow distinct evolutionary pathways. We emphasize that this conclusion is statistical in nature and does not exclude the possibility that individual galaxies may follow different evolutionary routes.

\item Our results disfavor AGN feedback as the dominant driver of rapid, galaxy-wide quenching in the local universe. Instead, AGN-PSBs are consistent with a scenario in which gas inflows trigger both central star formation and black hole activity, with the observed AGN activity representing a by-product rather than the primary cause of the PSB phase.

\end{enumerate}

\section*{acknowledgments}

Y.M.C. acknowledges support from the National Natural Science Foundation of China with grant No. 12333002 and the China Manned Space Project with No. CMS-CSST-2021-A08, and the Fundamental Research Funds for the Central Universities, No. KG202502. VW acknowledges Science and Technologies Facilities Council (STFC) grant ST/Y00275X/1, and Leverhulme Research Fellowship RF-2024-589/4. HL acknowledges support from a UKRI Frontier Research Grantee Grant (grant reference EP/Y037065/1). Funding for the Sloan Digital Sky Survey IV has been provided by the Alfred P. Sloan Foundation, the U.S. Department of Energy Office of Science, and the Participating Institutions. SDSS-IV acknowledges support and resources from the Center for High Performance Computing  at the University of Utah. The SDSS website is www.sdss.org. SDSS-IV is managed by the Astrophysical Research Consortium for the Participating Institutions of the SDSS Collaboration including the Brazilian Participation Group, the Carnegie Institution for Science, Carnegie Mellon University, Center for Astrophysics | Harvard \& Smithsonian, the Chilean Participation Group, the French Participation Group, Instituto de Astrof\'isica de Canarias, The Johns Hopkins University, Kavli Institute for the Physics and Mathematics of the Universe (IPMU) / University of Tokyo, the Korean Participation Group, Lawrence Berkeley National Laboratory, Leibniz Institut f\"ur Astrophysik Potsdam (AIP),  Max-Planck-Institut f\"ur Astronomie (MPIA Heidelberg), Max-Planck-Institut f\"ur Astrophysik (MPA Garching), Max-Planck-Institut f\"ur Extraterrestrische Physik (MPE), National Astronomical Observatories of China, New Mexico State University, New York University, University of Notre Dame, Observat\'ario Nacional / MCTI, The Ohio State University, Pennsylvania State University, Shanghai Astronomical Observatory, United Kingdom Participation Group, Universidad Nacional Aut\'onoma de M\'exico, University of Arizona, University of Colorado Boulder, University of Oxford, University of Portsmouth, University of Utah, University of Virginia, University of Washington, University of Wisconsin, Vanderbilt University, and Yale University.

\bibliography{ref}

\appendix

\begin{longtable}{lcccccccccc}
\caption{\raggedright The sample of AGN-PSB galaxies. (1) MaNGA identifier; (2) Right Ascension; (3) Declination; (4) inclination for galaxies; (5) redshift; (6) log $M_*$ fitted from K-corrected Sersic fluxes in the MaNGA DRP catalog, adjusted for $h = 0.7$; (7) S$\acute{\rm e}$rsic index; (8) D$_n$4000; (9) $g-r$ color index; (10) kinematic misalignment between gas and stars ($ |\rm{PA_{star}}- \rm{PA_{gas}}|$); (11) notes of unusual features (EML = Emission line). The full table is available as supplementary online material.}\label{tab:agnpsb}\\

\hline
MaNGAID & RA & Dec & $i$ & $z$ & log $M_*$ & $n$ & D$_n$4000 & $g-r$ & $\Delta$PA & Features \\
{(1)} & {(2)} & {(3)} & {(4)} & {(5)} & {(6)} & {(7)} & {(8)} & {(9)} & {(10)} & {(11)}\\
\hline
\endfirsthead
\hline
MaNGAID & RA & Dec & $i$ & $z$ & log $M_*$ & $n$ & D$_n$4000 & $g-r$ & $\Delta$PA & Features \\
{(1)} & {(2)} & {(3)} & {(4)} & {(5)} & {(6)} & {(7)} & {(8)} & {(9)} & {(10)} & {(11)}\\
\hline
\endhead
1-122304 & 121.92081 & 39.00424 & 26.1 & 0.02335 & 10.50 & - & 1.36 & 0.56 & 51.0 & misaligned \\
1-152306 & 196.79079 & 52.54641 & 52.1 & 0.02658 & 10.34 & 4.5 & 1.60 & 0.73 & 16.5 & - \\
1-163966 & 120.08742 & 26.61353 & 43.2 & 0.02674 & 11.02 & 3.7 & 1.52 & 0.61 & 5.0 & - \\
1-167688 & 155.88556 & 46.05775 & 29.9 & 0.02577 & 10.00 & - & 1.53 & 0.64 & 24.0 & - \\
1-196091 & 199.81627 & 54.60174 & 27.3 & 0.03237 & 10.74 & 4.1 & 1.63 & 0.68 & 2.0 & - \\
\multicolumn{11}{c}{$\cdots$} \\
\hline
\end{longtable}

\begin{longtable}{lcccccccccc}
\caption{\raggedright The sample of CPSB galaxies. Column descriptions are the same as in Table~\ref{tab:agnpsb}.}\label{tab:cpsb}\\

\hline
MaNGAID & RA & Dec & $i$ & $z$ & log $M_*$ & $n$ & D$_n$4000 & $g-r$ & $\Delta$PA & Features \\
{(1)} & {(2)} & {(3)} & {(4)} & {(5)} & {(6)} & {(7)} & {(8)} & {(9)} & {(10)} & {(11)}\\
\hline
\endfirsthead
\hline
MaNGAID & RA & Dec & $i$ & $z$ & log $M_*$ & $n$ & D$_n$4000 & $g-r$ KC003 & $\Delta$PA & Features \\
{(1)} & {(2)} & {(3)} & {(4)} & {(5)} & {(6)} & {(7)} & {(8)} & {(9)} & {(10)} & {(11)}\\
\hline
\endhead
1-115066 & 330.97675 & 12.66425 & 55.8 & 0.02677 & 9.69 & 2.3 & 1.59 & 0.60 & 1.5 & tidal tail \\
1-116828 & 342.78925 & 14.76518 & 36.6 & 0.09103 & 11.27 & 4.0 & 1.66 & 0.61 & 13.5 & tidal tail \\
1-121566 & 118.60759 & 35.07287 & 53.2 & 0.01415 & 9.05 & - & 1.22 & 0.53 & 56 & - \\
1-131060 & 224.54691 & 57.00097 & 34.5 & 0.02822 & 9.80 & 1.0 & 1.59 & 0.61 & 150 & misaligned \\
1-134964 & 246.76070 & 43.47610 & 31.7 & 0.04623 & 11.03 & 3.2 & 1.54 & 0.70 & 18.5 & tidal tail \\
\multicolumn{11}{c}{$\cdots$} \\
\hline
\end{longtable}

\begin{longtable}{lcccccccccc}
\caption{\raggedright The sample of RPSB galaxies. Column descriptions are the same as in Table~\ref{tab:agnpsb}.}\label{tab:rpsb}\\

\hline
MaNGAID & RA & Dec & $i$ & $z$ & log $M_*$ & $n$ & D$_n$4000 & $g-r$ & $\Delta$PA & Features \\
{(1)} & {(2)} & {(3)} & {(4)} & {(5)} & {(6)} & {(7)} & {(8)} & {(9)} & {(10)} & {(11)}\\
\hline
\endfirsthead
\hline
MaNGAID & RA & Dec & $i$ & $z$ & log $M_*$ & $n$ & D$_n$4000 & $g-r$ KC003 & $\Delta$PA & Features \\
{(1)} & {(2)} & {(3)} & {(4)} & {(5)} & {(6)} & {(7)} & {(8)} & {(9)} & {(10)} & {(11)}\\
\hline
\endhead
1-106664 & 347.26196 & 0.26696 & 62.3 & 0.03251 & 10.32 & 1.9 & 1.34 & 0.56 & 27.0 & - \\
1-122139 & 122.32324 & 38.33694 & 43.8 & 0.04072 & 10.00 & 2.7 & 1.34 & 0.56 & 11.0 & - \\
1-138106 & 144.35231 & 48.51545 & 36.6 & 0.02435 & 10.24 & 3.5 & 1.27 & 0.43 & 9.5 & - \\
1-149557 & 171.77902 & 51.13165 & 33.2 & 0.01450 & 9.20 & 1.4 & 1.42 & 0.52 & 155.5 & PA$_{\rm err}>60^{\circ}$ \\
1-152828 & 117.92943 & 30.44875 & 51.1 & 0.01424 & 8.96 & 1.9 & 1.19 & 0.29 & 10.5 & tidal tail \\
\multicolumn{11}{c}{$\cdots$} \\
\hline
\end{longtable}
    
\end{document}